\documentclass[twocolumn]{aastex61}

\shorttitle{Massive quiescent galaxies at $z\sim4$}
\shortauthors{Kubo et al.}
\definecolor{grey}{rgb}{0.6, 0.6, 0.6}

\begin{document}

\title{The rest-frame optical sizes of massive galaxies with suppressed star formation at $z\sim4$ }

\email{mariko.kubo@nao.ac.jp}

\author[0000-0000-00000-0000]{Mariko Kubo}
\affiliation{National Astronomical Observatory of Japan 
2-21-1, Osawa, Mitaka, Tokyo, 181-8588, Japan}

\author{Masayuki Tanaka}
\affiliation{National Astronomical Observatory of Japan 
2-21-1, Osawa, Mitaka, Tokyo, 181-8588, Japan}

\author{Kiyoto Yabe}
\affiliation{Kavli Institute for the Physics and Mathematics of the Universe, The University of Tokyo, 5-1-5 Kashiwanoha, Kashiwa, Chiba 277-8583, Japan}

\author{Sune Toft}
\affiliation{Dark Cosmology Centre, Niels Bohr Institute, University of Copenhagen, Juliane Maries Vej 30, DK-2100 Copenhagen, Denmark}
\affiliation{Cosmic Dawn Center (DAWN), Niels Bohr Institute, University of Copenhagen, Copenhagen, Denmark}

\author{Mikkel Stockmann}
\affiliation{Dark Cosmology Centre, Niels Bohr Institute, University of Copenhagen, Juliane Maries Vej 30, DK-2100 Copenhagen, Denmark}
\affiliation{Cosmic Dawn Center (DAWN), Niels Bohr Institute, University of Copenhagen, Copenhagen, Denmark}

\author{Carlos G\'omez-Guijarro}
\affiliation{Dark Cosmology Centre, Niels Bohr Institute, University of Copenhagen, Juliane Maries Vej 30, DK-2100 Copenhagen, Denmark}
\affiliation{Cosmic Dawn Center (DAWN), Niels Bohr Institute, University of Copenhagen, Copenhagen, Denmark}



\begin{abstract}
We present the rest-frame optical sizes of massive quiescent galaxies (QGs) at $z\sim4$ 
measured at $K'$-band with the Infrared Camera and Spectrograph (IRCS) 
and adaptive optics facility, AO188, on the Subaru telescope.
Based on a deep multi-wavelength catalog in the Subaru XMM-Newton Deep Survey Field (SXDS),
covering a wide wavelength range from the $u$-band to the IRAC $8.0\mu m$ over 0.7 deg$^2$,
we evaluate photometric redshift to identify massive ($M_{\star}\sim10^{11}\ M_\odot$) 
galaxies with suppressed star formation.
These galaxies show a prominent Balmer break feature at $z\sim4$, suggestive of an evolved
stellar population.  
We then conduct follow-up $K'$-band imaging with adaptive optics 
for the five brightest galaxies ($K_{AB,total}=22.5\sim23.4$).
Compared to lower redshift ones, QGs at $z\sim4$ have smaller physical sizes 
of effective radii $r_{eff}=0.2$ to $1.7$ kpc.
The mean size measured by stacking the four brightest objects, 
a more robust measurement,  is $r_{eff}=0.5\rm\ kpc$.
This is the first measurement of the rest-frame optical sizes of QGs at $z\sim4$. 
We evaluate the robustness of our size measurements using simulations and find that our size estimates
are reasonably accurate with an expected systematic bias of $\sim0.2$ kpc. 
If we account for the stellar mass evolution, massive QGs at $z\sim4$ 
are likely to evolve into the most massive galaxies today. 
We find their size evolution with cosmic time 
in a form of $\log(r_e/{\rm kpc})= -0.56+1.91 \log(t/\rm Gyr)$. 
Their size growth is proportional to the square of stellar mass, 
indicating the size-stellar mass growth driven by minor dry mergers. 
\end{abstract}




%----------------------------------------------
\section{Introduction} \label{sec:intro}

There is mounting evidence for the presence of massive galaxies with suppressed star formation
at $z>2$ (e.g., \citealt{2005ApJ...626..680D, 2008ApJ...677L...5V}).
These galaxies are known to be remarkably compact and dense compared to local ones 
(e.g., \citealt{2006MNRAS.373L..36T,2007ApJ...671..285T, 2008ApJ...677L...5V, 2014ApJ...788...28V,2017MNRAS.469.2235K}). 
The size evolution of these massive quiescent galaxies (QGs) can be parameterized as
$r\propto(1+z)^{\beta}$ where $\beta\sim-1.5$  which is steeper than $\beta \sim -1$ of
star forming galaxies (SFGs; e.g., \citealt{2014ApJ...788...28V,2015ApJS..219...15S}). 
%$\beta=-0.7\sim-1.2$

The remarkable compactness and early formation of massive QGs 
pose a challenge to the standard picture of galaxy formation 
in which galaxies grow hierarchically and become more massive with time. 
Gas rich major mergers (e.g., \citealt{2008ApJS..175..356H,2015MNRAS.449..361W}) 
and infall of giant clumps formed via disk instability 
(e.g., \citealt{2008ApJ...688...67E, 2009ApJ...703..785D}) 
can trigger nuclear starburst and increase 
the central density in galaxies to form a compact remnant. 
Discoveries of compact starburst galaxies at $z>2$ may support these scenarios
\citep{2014ApJ...782...68T, 2014ApJ...791...52B, 2015ApJ...810..133I,2016ApJ...827L..32B, 2017ApJ...849L..36I, 2017ApJ...840...47B,2018ApJ...856..121G}. 
On the other hand, massive QGs at high redshift
need several to ten times growth in size but less growth in stellar mass 
to evolve into giant elliptical galaxies today. 
Dry minor mergers (e.g., \citealt{2009ApJ...697.1290B,2009ApJ...699L.178N}), 
adiabatic expansion \citep{2008ApJ...689L.101F, 2014ApJ...791...45V}
and size evolution of newly quenched galaxies 
with redshift \citep{2013ApJ...773..112C, 2013ApJ...777..125P, 2015ApJ...799..206B} 
have been proposed as the driver of this steep size growth. 

Now massive QGs at $z\sim4$ 
are found photometrically \citep{2014ApJ...783L..14S} 
and confirmed spectroscopically 
($z_{spec}=3.717$; \citealt{2017Natur.544...71G, 2018A&A...611A..22S}).
The {\it Hubble Space Telescope (HST)} has been the main workhorse 
in the field of galaxy morphologies at high redshift, but
it can not probe the rest-frame optical wavelength regime of galaxies at $z>3$ 
due to its wavelength cutoff of $\sim1.7~\mu m$.
In this study, we select galaxies with a prominent 
Balmer break feature at $z\sim4$ photometrically 
from the Subaru XMM-Newton Deep Survey (SXDF; \citealt{2008ApJS..176....1F})
and investigate their rest-frame optical morphologies by the deep $K'$-band images obtained with
the adaptive optics (AO) on the Subaru Telescope. 

This paper is organized as follows: 
in Section 2 we describe our sample selection of massive galaxies with suppressed star formation, 
in Section 3 we describe the observation and data reduction procedure, 
in Section 4 we describe the size measurement method and possible errors, 
and in Section 5 we show the results. 
We discuss the stellar mass surface density and size-stellar mass evolution of them in Section 6. 
Throughout the paper, we adopt a $\Lambda$CDM cosmology with $H_0=70$ km s$^{-1}$ Mpc$^{-1}$, 
$\Omega_{\Lambda} = 0.7$ and $\Omega_m = 0.3$, 
and magnitudes are given in the AB system.

%% Note that the \setcounter and \renewcommand are needed here because
%% this example is using a mix of deluxetable and tabular.  Here the
%% deluxetable counters are set with \tablenum but the situation is a bit
%% more complex for tabular.  Use the first command to set the Table number
%% to ONE LESS than it should be.  The next command will auto increment it
%% to the desired number.
\section{Sample construction}
\label{sec:sample_construction}
\subsection{Multi-band Catalog}

We base our analysis on a multi-band photometric catalog in the Subaru
XMM-Newton Deep Field (SXDF; \citealt{2008ApJS..176....1F}).
SXDF has deep optical imaging from Suprime-Cam of the Subaru Telescope 
in $BVRiz$-bands \citep{2008ApJS..176....1F}.
The UKIRT Infrared Deep Sky Survey (UKIDSS;  \citealt{2007MNRAS.379.1599L}) 
is centered on the same field and we use the Data Release
10 to complement the optical data.
Furthermore, the $u$-band photometry from CFHT Megacam and {\it Spitzer} photometry
from the {\it Spitzer} UKIDSS Ultra Deep Survey (SpUDS; \citealt{2007sptz.prop40021D}) 
are available, allowing us to cover the entire optical
and IR wavelengths up to $24\mu m$ over a wide area.  It is an excellent
field to search for faint, rare objects at high redshifts.

We first register all the optical images to the WCS grid of the UKIDSS images.
The seeing is different from band to band, and we apply  a Gaussian kernel
to homogenize the seeing to $\sim0.82$ arcsec.
We run {\sf SExtractor} \citep{1996A&AS..117..393B} on the $K$-band image to detect
sources.  We then measure sources in the other optical and nearIR bands
using the dual image mode.
We perform photometry within a circular aperture of 2.0 arcsec in all the bands.
Because we miss a fraction of total light in this aperture, we measure the Kron
fluxes of objects in the $K$-band and estimate the aperture correction, assuming
the Kron flux is the total flux  
(here after we refer to the Kron magnitude as the total magnitude). 
We apply the aperture correction to the 2.0 arcsec
aperture photometry in all the bands so that our photometry is closer to total
light while keeping the accurate colors.

Because of the relatively large PSF sizes of the {\it Spitzer}/IRAC images,
objects are often blended with nearby objects, and we choose to perform
the {\it Spitzer} photometry separately from the optical-nearIR bands.
We use {\sf T-PHOT} \citep{2015A&AS..582..15M} version 1.5.11 to fit 2d profiles
of objects in the IRAC images taking the object blending into account using
the $K$-band image as a prior.
For objects detected in the $K$-band high-resolution image (HRI), small image cutouts
of the same region are generated in order to model the IRAC low-resolution image (LRI).
The cutouts are convolved with a kernel constructed from LRI and HRI, both of
which are constructed from point sources selected in HRI, to homogenize the PSF.　
Then, the optimization process is performed by scaling the fluxes of the objects
of the PSF-matched HRI to match the LRI using the $\chi^{2}$ minimization technique.
We process the IRAC images in all channels from 3.6$\mu$m to 8.0$\mu$m in the same way,
and we use the total magnitude of each object from the best-fit model flux.

In the final catalog, we have about $10^5$ objects over $\sim0.7$ deg$^2$
with coverage in all the filters.  Table \ref{tab:depth} summarizes the depth
in each band.

%----------------------
\begin{table}
 \centering
 \caption{
   $5\sigma$ limiting magnitudes within 2 arcsec apertures for each filter.
 }
 \begin{tabular}{ccc}
   filter & instrument & depth\\
   \hline
   $u$ & Megacam      & 26.8\\
   $B$ & Suprime-Cam  & 27.6\\
   $V$ & Suprime-Cam  & 27.3\\
   $R$ & Suprime-Cam  & 27.1\\
   $i$ & Suprime-Cam  & 27.0\\
   $z$ & Suprime-Cam  & 26.0\\
   $J$ & WFCAM        & 25.2\\
   $H$ & WFCAM        & 24.6\\
   $K$ & WFCAM        & 25.0\\
   ch1 & IRAC         & 24.8\\
   ch2 & IRAC         & 24.3\\
   ch3 & IRAC         & 22.6\\
   ch4 & IRAC         & 22.5\\
   \hline
   \label{tab:depth}
 \end{tabular}
\end{table}

%----------------------------------------------
\subsection{Target Selection}

We run a custom photometric redshift code \citep{2015ApJ...801...20T}
on the multi-band catalog.
 This is a template-fitting code and we use templates generated using
 the \citet{2003MNRAS.344.1000B} stellar population synthesis code.
  We adopt the following assumptions in the models: exponentially declining
 star formation history, solar metallicity, \citet{1994ApJ...429..582C}
 attenuation curve, and \citet{2003PASP..115..763C} initial mass function (IMF).
As we know the SFR and attenuation of each template, we add emission lines due to star formation using the emission line intensity ratios by \citet{2011MNRAS.415.2920I} (see \citealt{2015ApJ...801...20T} for details).
The code infers redshifts and physical
properties of galaxies such as stellar mass in a self-consistent manner
 and the uncertainties on the physical properties quoted in the paper have been
 estimated by marginalizing over all the other parameters, including redshift.
As we have a large number of filters spanning a wide wavelength range, the data
has a strong constraining power on the overall SED shapes.  We therefore
choose to apply flat priors in the fitting.  We have confirmed that our results do not signifiantly change if we apply the full priors.
Using some of the publicly available spectroscopic redshifts
(\citealt{2013MNRAS.433..194B,2013MNRAS.428.1088M}, Simpson et al. in prep),
we achieve a normalized dispersion of $\sigma(\Delta z/(1+z))=0.029$ and
an outlier rate of 4.8\%, where the outliers are defined in the conventional
way (i.e., those with $|\Delta z/(1+z)|>0.15$; \citealt{2017arXiv170405988T}).  However,
the spectroscopic sample is heterogeneous and the numbers here should not
be over-interpreted.

\begin{figure}
\centering
\includegraphics[width=80mm]{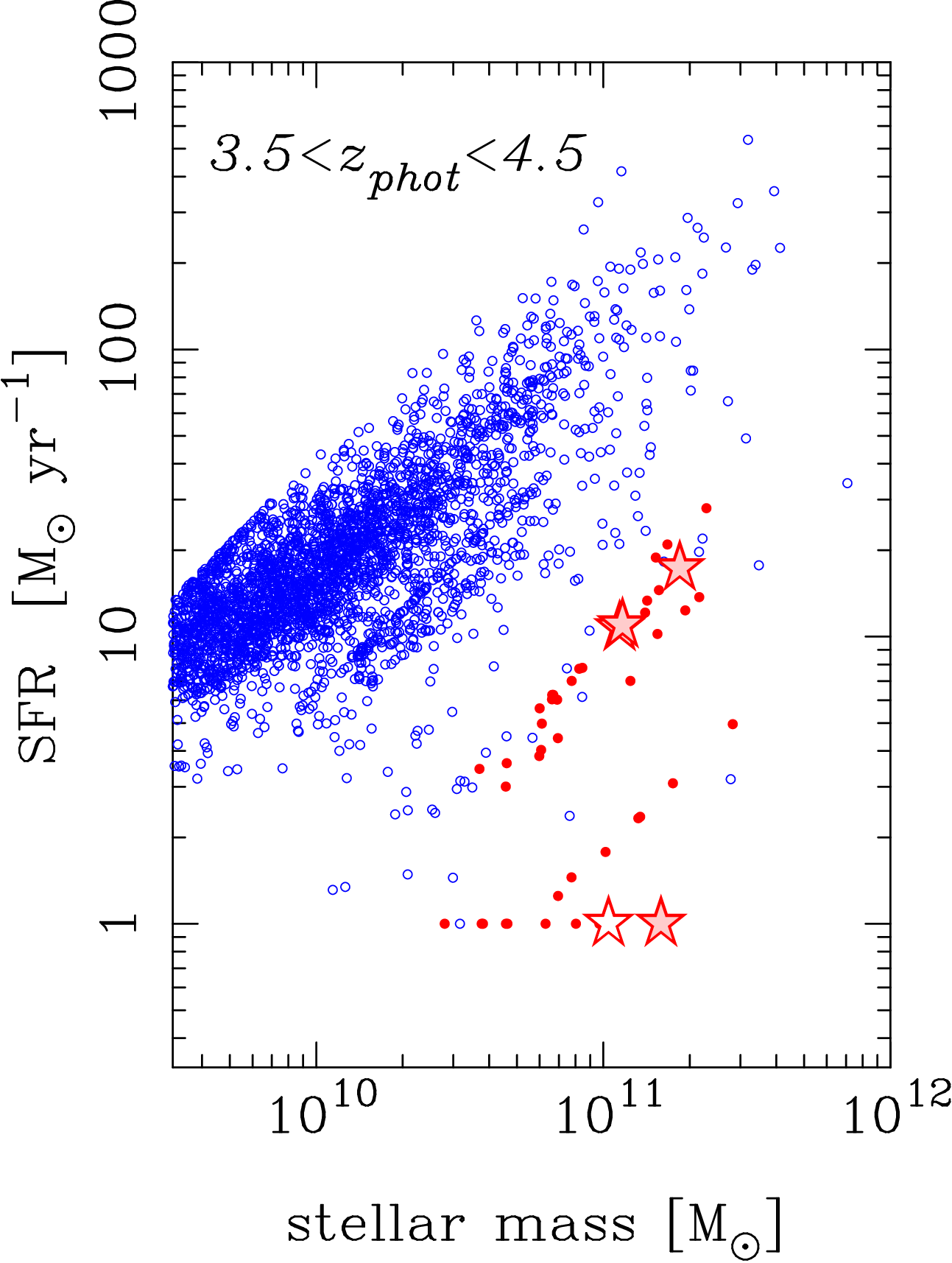}
\caption{
  SFR v.s. stellar mass of galaxies at $z\sim4$.
  The open circles are SFGs.
  The filled circles are QG candidates with a $1\sigma$ upper limit
  of the sSFR lower than $10^{-9.5}\rm\ yr^{-1}$.
  Objects with SFRs smaller than 1$\rm M_\odot\ yr^{-1}$ are shown at SFR=$1\rm M_\odot\ yr^{-1}$
  only for illustrative purposes.
  The stars indicate the targets observed with IRCS+AO188 (see \S
  \ref{sec:observation_and_data_reduction}). 
  The open star is ID3, the target 
  also classified as quiescent by \citet{2014ApJ...783L..14S}.
}
\label{fig:sfr_smass}
\end{figure}

We exclude objects with unreliable photo-$z$'s using the reduced chi-squares,
$\chi_\nu>4$.  Poor chi-squares are often due to poor photometry (e.g., halos
around bright stars and object blending).  For the purpose of this paper,
we do not need a complete sample of evolved galaxies at high redshift and
this cut does not introduce any bias.
We then select galaxies at $3.5<z_{phot}<4.5$.  Fig. \ref{fig:sfr_smass} shows
star formation rate (SFR) against stellar mass of the $z\sim4$ galaxies.
Both SFR and stellar mass are from the SED fit.  There is a clear sequence of
SFGs and also a population of massive galaxies with suppressed
star formation.
These two populations can be separated very well at specific SFR (sSFR)
of $10^{-9.5}\rm yr^{-1}$.  
To be conservative, we choose galaxies whose $1\sigma$ upper limit 
of their sSFR is lower than $10^{-9.5}\rm yr^{-1}$
as the targets for the near-IR follow-up imaging with AO. 
The red filled points in Fig. \ref{fig:sfr_smass} satisfy this condition.
We note that there is some ambiguity in the definition of QGs
in the literature, but when we refer to QGs in what follows,
we mean galaxies with suppressed star formation as defined in Fig.~\ref{fig:sfr_smass}.
The $UVJ$ diagram is often used to define QGs, but it is tuned at $z\lesssim2$
\citep{2005ApJ...624L..81L,2009ApJ...691.1879W} and is not clear whether it can be applied
to $z\sim4$ galaxies.  For this reason, we adopt the sSFR-based definition.

\begin{figure*}
\centering
\includegraphics[width=80mm]{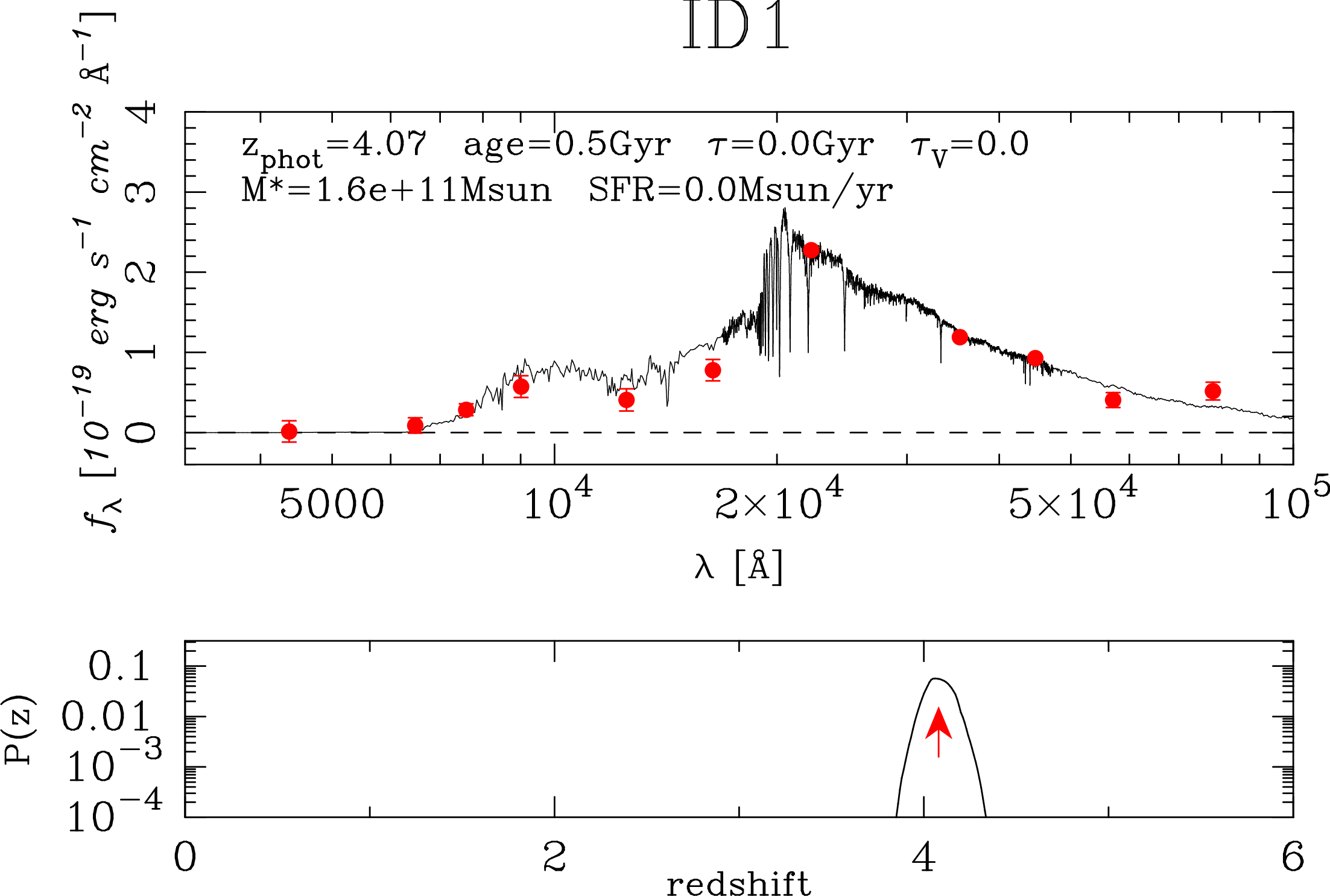}
\includegraphics[width=80mm]{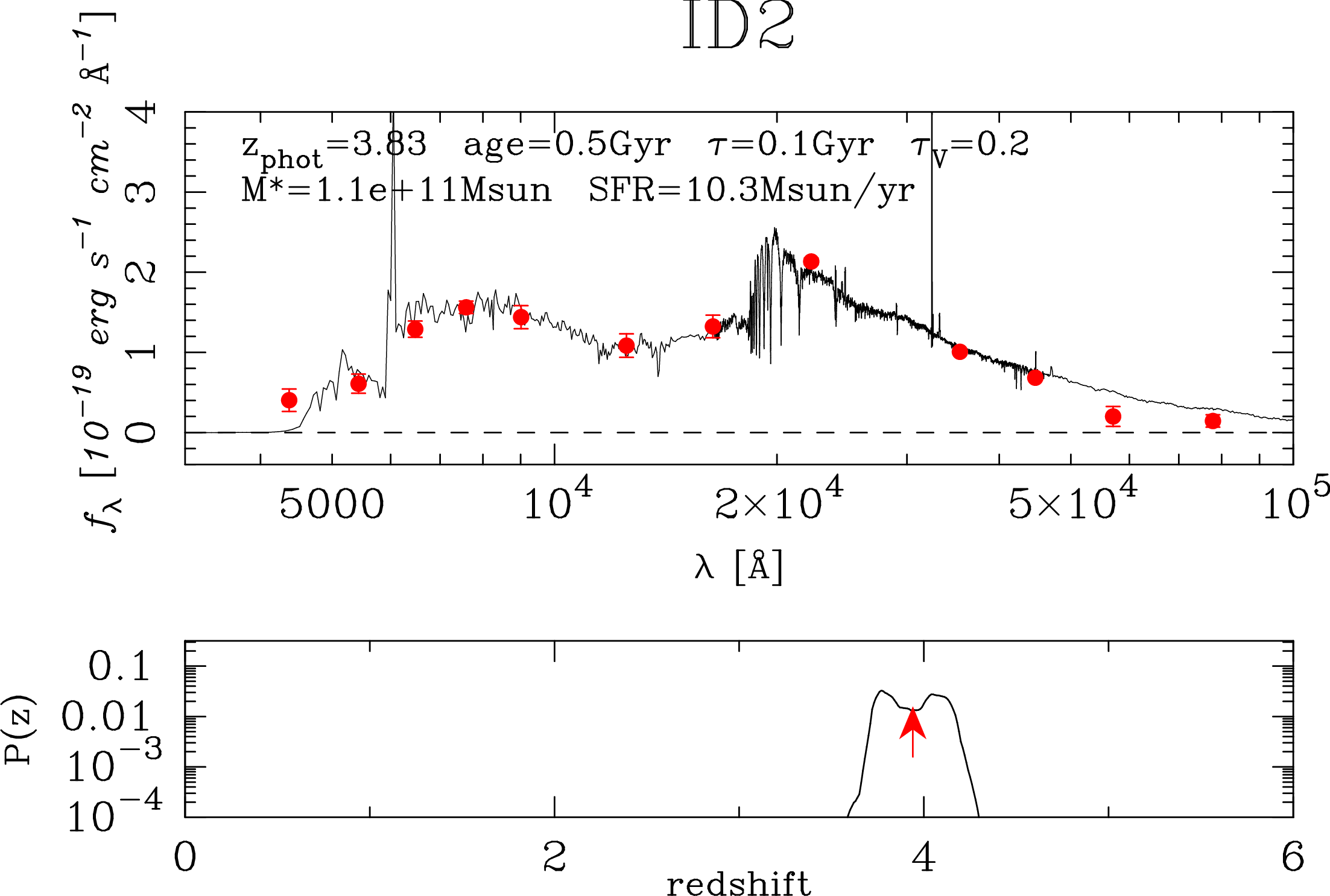}\\\vspace{0.5cm}
\includegraphics[width=80mm]{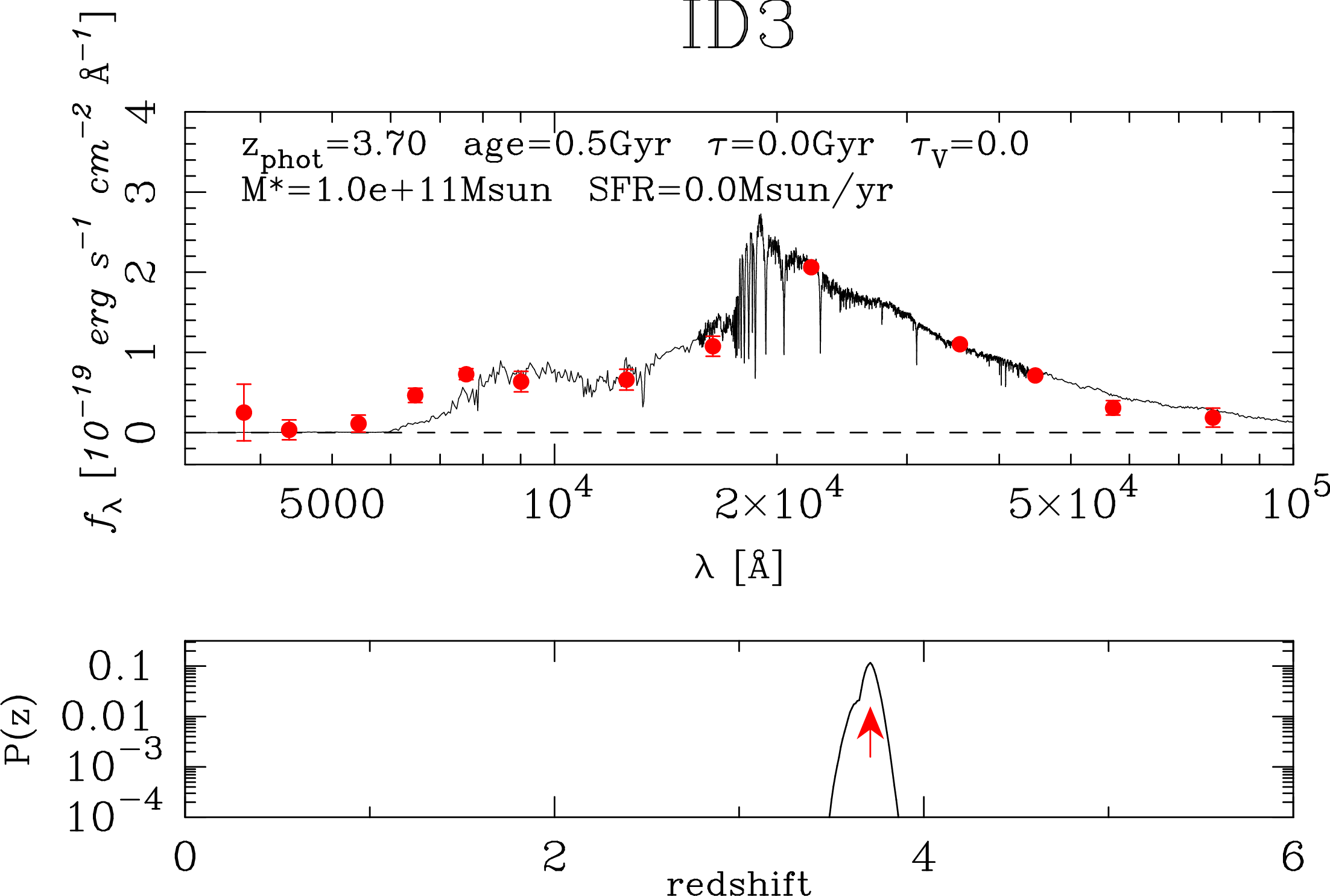}
\includegraphics[width=80mm]{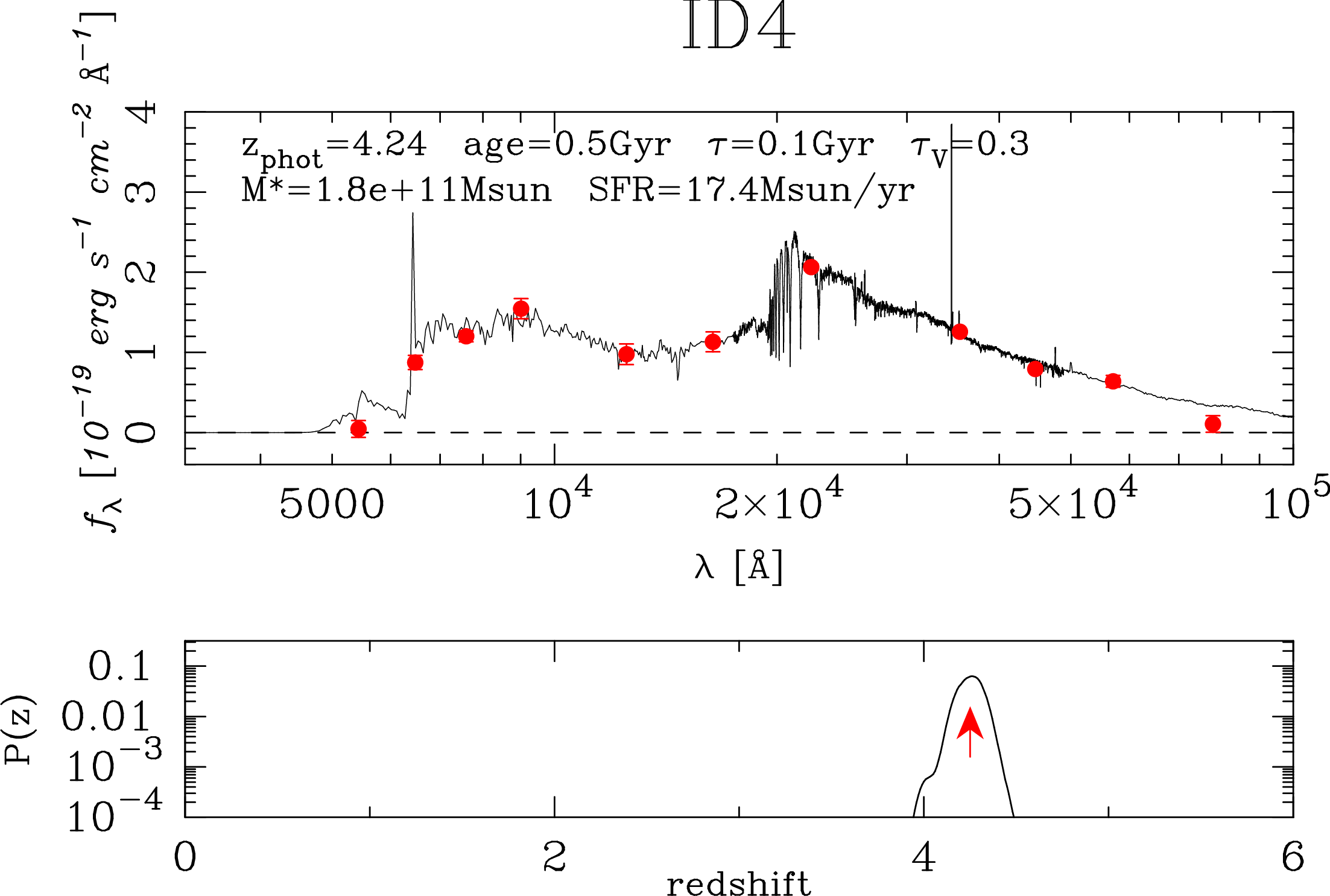}\\\vspace{0.5cm}
\includegraphics[width=80mm]{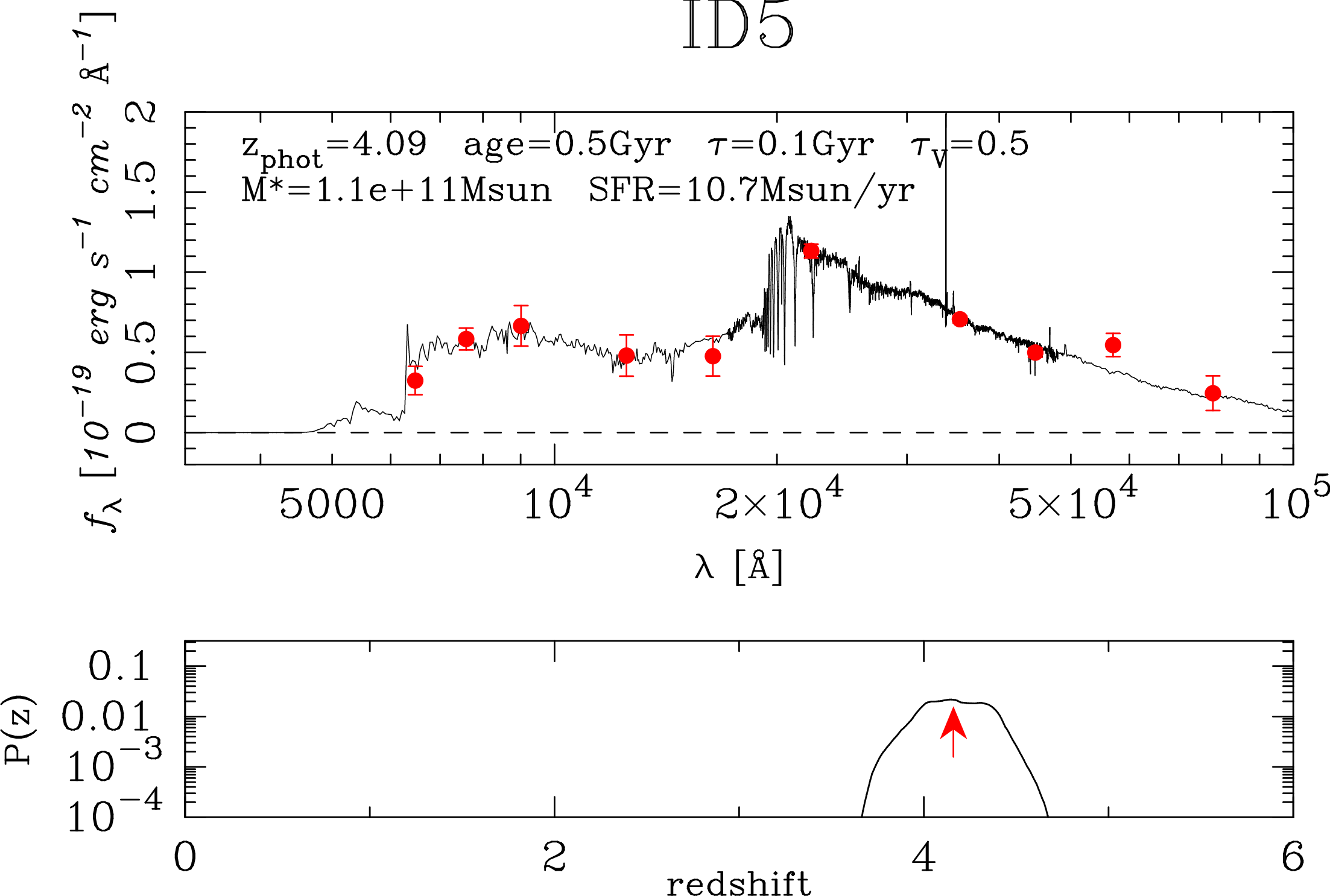}
\caption{
  {\it Top:}
  The SEDs of our targets, ID1 to ID5.  The spectrum is the best-fitting template spectrum
  and the points are the observed photometry.  Some of the relevant quantities such as
  age and star formation timescale of the template are also indicated.
  {\it Bottom:}
  Redshift probability distribution function.  The arrow shows the median redshift.
}
\label{fig:individual_seds}
\end{figure*}

\begin{figure}
\centering
\includegraphics[width=80mm]{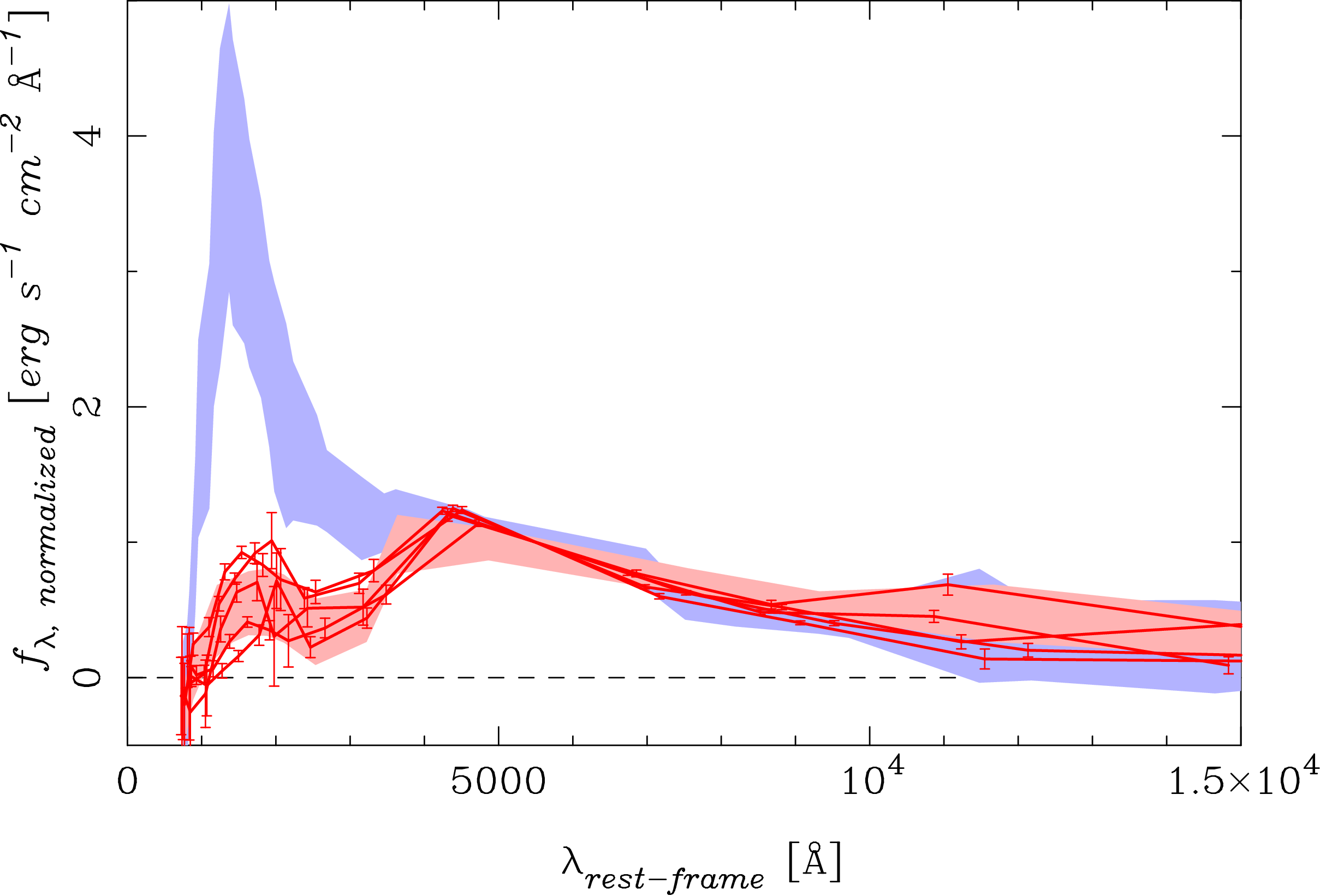}
\caption{
  Rest-frame mean SEDs of SFGs (blue) and QGs (red) normalized in
  the $V$-band.  The shaded areas encompass the 68 percentile of the distribution.
  The objects that we observed are shown as the solid lines.
  They show the prominent Balmer break.
}
\label{fig:mean_seds}
\end{figure}

In addition to the sSFR constraint, further practical constraints come 
from the location of tip-tilt stars for the AO-assisted observation. 
Since we need tip-tilt stars $R=16.5$ or brighter (for NGS mode, $R<19$ for LGS mode) for AO188, 
the available targets are further limited. 
We have conducted the near-IR follow-up imaging with AO 
for five of the brightest QGs with suitable tip-tilt stars as shown by the stars
in Fig. \ref{fig:sfr_smass} (here after ID1-5).
Fig. \ref{fig:individual_seds} shows their SEDs.  
All of them are located around $z_{phot}\sim4$.  
As can be seen, all the objects show a prominent
Balmer break, indicative of an evolved stellar population.  
ID1 and ID2 have a faint UV continuum and are consistent with passively evolving galaxies
(SED-based SFR is consistent with zero).
The others have a brighter UV continuum, but the break feature is still prominent.
To further characterize our targets,
we compare the mean SEDs of SFGs 
with that of QGs in Fig. \ref{fig:mean_seds}.  
SFGs have a very blue UV continuum with a strong Lyman break.
On the other hand, the SEDs of our targets are clearly distinct; they have a suppressed
UV continuum with a clear Balmer break.  This break cannot be due to dust extinction
because it does not introduce a sudden break at 3650\AA\ while keeping the continuum
at longer wavelengths blue.  This is due to abundant A-type stars in these galaxies.
The observed targets are consistent with the mean SED shown by the red shades and that
suggests that they are representative of the evolved population around that redshift.

We note that a part of our survey area is observed 
in the Fourstar Galaxy Evolution survey (ZFOURGE) \citep{2016ApJ...830...51S}.
\citet{2014ApJ...783L..14S} select QGs at $z\sim4$ using
the rest-frame $UVJ$ colors and photometric redshifts from ZFOURGE.
We briefly compare our sample of QGs with those in \citet{2014ApJ...783L..14S}.
We find that the QGs identified in SXDF (UDS) in \citet{2014ApJ...783L..14S} all satisfy
sSFR$<10^{-9.5}\rm yr^{-1}$ based on our catalog.
On the other hand, two of our targets, ID3 and ID5 are in the ZFOURGE field. 
ID3 is also identified as a QG in ZFOURGE, whereas ID5 is not.
The rest-frame color of ID5 is $U-V=0.95\pm0.04$ and $V-J=0.86\pm0.02$ \citep{2016ApJ...830...51S}, 
slightly bluer than the color criterion for QGs adopted in \citet{2014ApJ...783L..14S},
but their SED fit suggests sSFR $<10^{-9.5}\rm yr^{-1}$ at $z_{\rm phot}\approx4$, 
satisfying our criterion of QGs.
Overall, our QG selection is broadly compatible with that of \citet{2014ApJ...783L..14S}.
It is noteworthy that, most of QGs in their sample are fainter than $K>23$.
Thanks to the wider area coverage, most of our targets are brighter
and better suited for detailed structural studies.

%----------------------

%----------------------

%----------------------

%----------------------------------------------
\section{Observation and data reduction}
\label{sec:observation_and_data_reduction}

We observed the five targets selected in \S 2
with IRCS \citep{1998SPIE.3354..512T, 2000SPIE.4008.1056K}+AO188
\citep{2008SPIE.7015E..10H, 2010SPIE.7736E..0NH} on the Subaru Telescope 
on the 25th and 26th of September 2016.  We used the $K'$ filter with
the 52 mas pixel scale.  The observing conditions were fair; the sky
was clear on both nights with reasonably good seeing ($\sim0.2$ arcsec with AO),
though it fluctuated occasionally.  We observed both in NGS and
LGS modes due to occasional poor seeing and satellite crossings. 
We reject the worst $\sim10$\% of the bad seeing frames.
After rejecting these bad PSF frames, the variation of PSF sizes of the frames on
each target is less than 0.05 arcsec.

We reduced the data using the {\sf IRAF} data reduction tasks 
following the data reduction manual for
the IRCS\footnote{http://www.subarutelescope.org/Observing/DataReduction\\/Cookbooks/IRCSimg\_2010jan05.pdf}.
We first mask bad pixels and then apply the flat, which were constructed from dithered
science exposures with objects masked out.  
The sky background is the median value in the whole area 
of each frame, $\sim54$ arcsec on a side.  
We estimate the telescope offset between the pointing
from the relative positions of bright stars within the field of view.
Finally, we combine the frames with 3 sigma clipping.

Magnitude zero-points are calibrated by using the $K$-band images of UKIDSS. 
We estimate $K-K'$ (i.e., WFCAM - IRCS) color as a function of $J-K$ color using
the stellar library from \citet{1998PASP..110..863P}.
We set the zero points of the IRCS-AO $K'$-band images by matching the fluxes 
of bright ($K<21$) but not saturated stars 
with those measured on the UKIDSS $K$-band images after applying the $K-K'$ color term.
The $K-K'$ colors of the stars used as the standard stars here range from $-0.12$ to $-0.10$.  
Since observing conditions were stable during the nights, 
we use the average of magnitude zero-points of each night, 
$25.41$ for 25th Sep (ID2 \& 3) and $25.43$ for 26th Sep (ID1,4 \& 5).

We summarize the details of the coadd images in Table \ref{tab:targetdiscription}. 
The total exposure time of each target ranges from 18 to 54 minutes. 
The FWHM PSF sizes measured on the PSF reference stars range from $0.15$ to $0.23$ arcsec. 
%Here we measure the limiting magnitudes with $0.3$ arcsec diameter apertures. 

%--------------------------
\begin{deluxetable*}{lccccccc}
\tablecaption{Summary of observations \label{tab:targetdiscription}}
\tablecolumns{8}
\tablewidth{0pt}
\tablehead{
\colhead{ID} &
\colhead{R.A.} &
\colhead{Dec} & \colhead{EXPTIME} & \colhead{ZEROPOINT} &  \colhead{depth\tablenotemark{a}}&  \colhead{separation\tablenotemark{b}}  & \colhead{FWHM PSF\tablenotemark{c}}\\
\colhead{} & \colhead{(h:m:s)} & \colhead{(d:m:s)}  & \colhead{(min)} & \colhead{(mag)} & \colhead{(mag)} & \colhead{(arcsec)} & \colhead{(arcsec)}
}
\startdata
   1 &   02:19:01.511 &  -05:18:29.07 & 33 & $25.43$ &	24.7  &  72(33) & 0.17\\ %SXDS1(old ID)
   2 &   02:17:59.073 &  -05:09:39.89 & 18 & $25.43$ &	24.6 & 53(34) & 0.21\\ % SXDS5
   3 &   02:17:22.781 &  -05:17:33.34 & 35 & $25.41$ &	24.9 & 48(16) & 0.15\\ %SXDS2
   4 &   02:17:19.833 &  -04:43:34.75 & 43 & $25.43$ &	25.0 & 41(38) & 0.23\\ %SXDS8
   5 &   02:16:58.232 &  -05:08:35.21 & 54 & $25.41$ &	25.0 & 37(13) & 0.19\\ %SXDS6
\enddata
\tablenotetext{a}{5$\sigma$ limiting magnitudes measured with $0.3$ arcsec diameter aperture.}
\tablenotetext{b}{The separation between the tip-tilt stars and the targets. 
  The numbers in the parentheses are the separations between the tip-tilt stars and the PSF reference stars. }
\tablenotetext{c}{FWHM of the PSF reference stars.}
\end{deluxetable*}
%ID5 within UDS Fourstar+HST
%ID3 within UDS Fourstar but no HST

%----------------------------------------------
\section{Size measurement}
\subsection{Flux completeness}

We first examine the flux completeness of our targets on the IRCS-AO $K'$-band images 
by comparing the flux measured on the IRCS-AO $K'$-band and UKIDSS $K$-band images. 
The S/N on our $K'$-band images are lower than that on the UKIDSS $K$-band images. 
Then if our targets have morphologies dominated by low-surface brightness components, 
large fraction of their fluxes detectable on the UKIDSS $K$-band images 
may go below the detection limit on our IRCS-AO $K'$-band images. 
Also, the AO-corrected PSF tends to have an extended wing, which also
introduces a diffuse component in the observed profiles.  These effects can
result in underestimated sizes and fluxes.

\begin{figure}
\centering
\includegraphics[width=85mm]{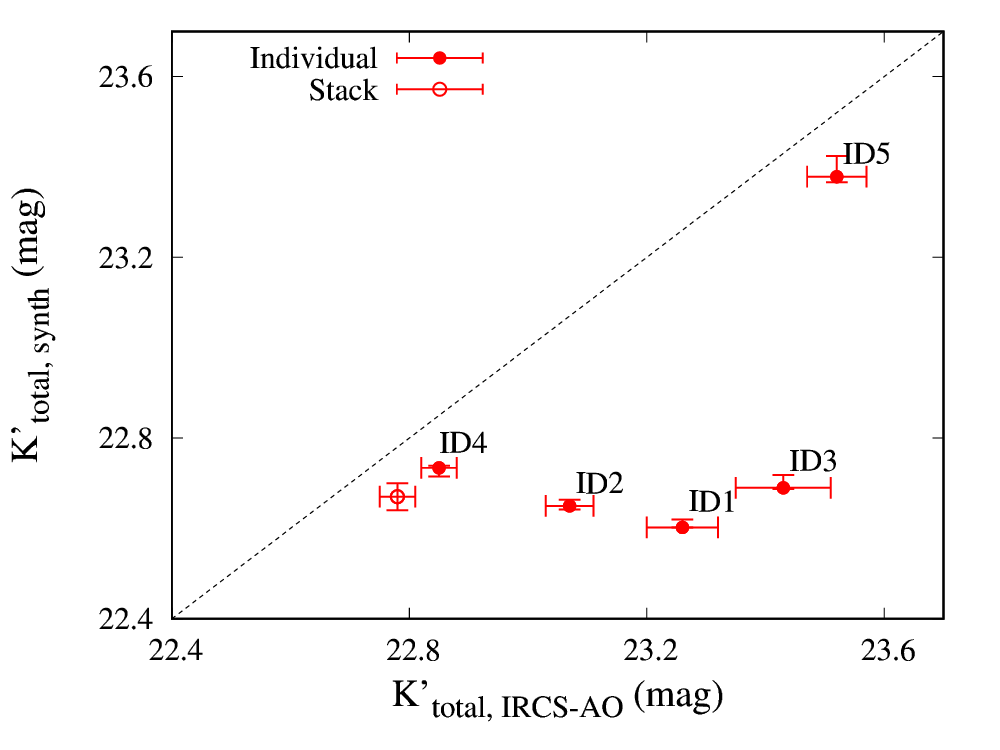}
\caption{
  Synthetic $K'$-band magnitude ($K'_{total, synth}$) plotted 
  against IRCS $K'$-band magnitude ($K'_{total, IRCS-AO}$) of our targets.
 The red filled circles show the individual objects. 
 The red open circle shows the stack of ID$1$ to ID$4$.
}
\label{fig:magauto}
\end{figure}

We compare the $K'$-band total magnitudes measured 
on our IRCS-AO $K'$-band ($K'_{total, IRCS-AO}$) 
with the UKIDSS $K$-band magnitudes corrected 
of the $K-K'$ color term using the best-fit SED model 
($K'_{total, synth}$) in order to evaluate the missing flux 
(Fig. \ref{fig:magauto} and Table \ref{tab:result}).
Overall, we tend to underestimate the fluxes in the IRCS-AO images as expected.
For ID4 and ID5, we underestimate only by 10\% and the missing light probably
does not affect our size measurements significantly.
However, we miss $25-50$\% of the light for the other targets.  
Though a care is needed when interpreting individual galaxies, the stacked galaxy 
(open circle, see \S 4.4)
does not show a significant amount of missing flux,
suggesting that its size can be robustly measured.
We make an attempt to estimate the effects 
of the missing light on the size measurements in Section~\ref{sec:galfit_fitting_errors},
where we actually reproduce the amount of the missing fluxes with a simulation and evaluate the limitation from PSF.

%--------------------------

\begin{figure}
\centering
\includegraphics[width=85mm]{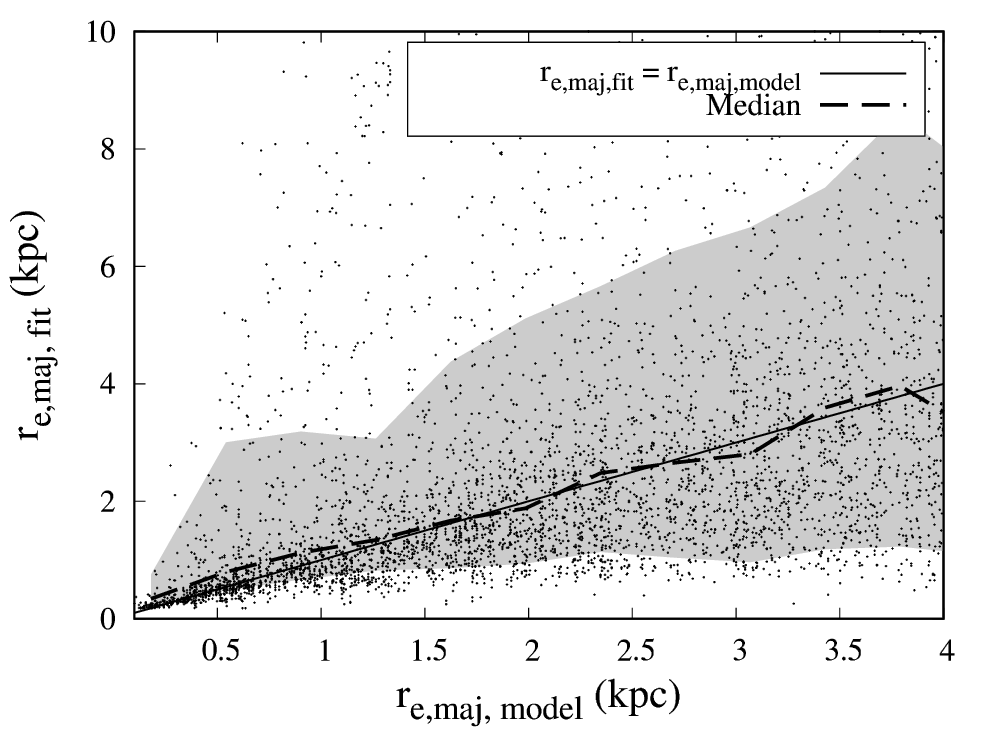}
\includegraphics[width=85mm]{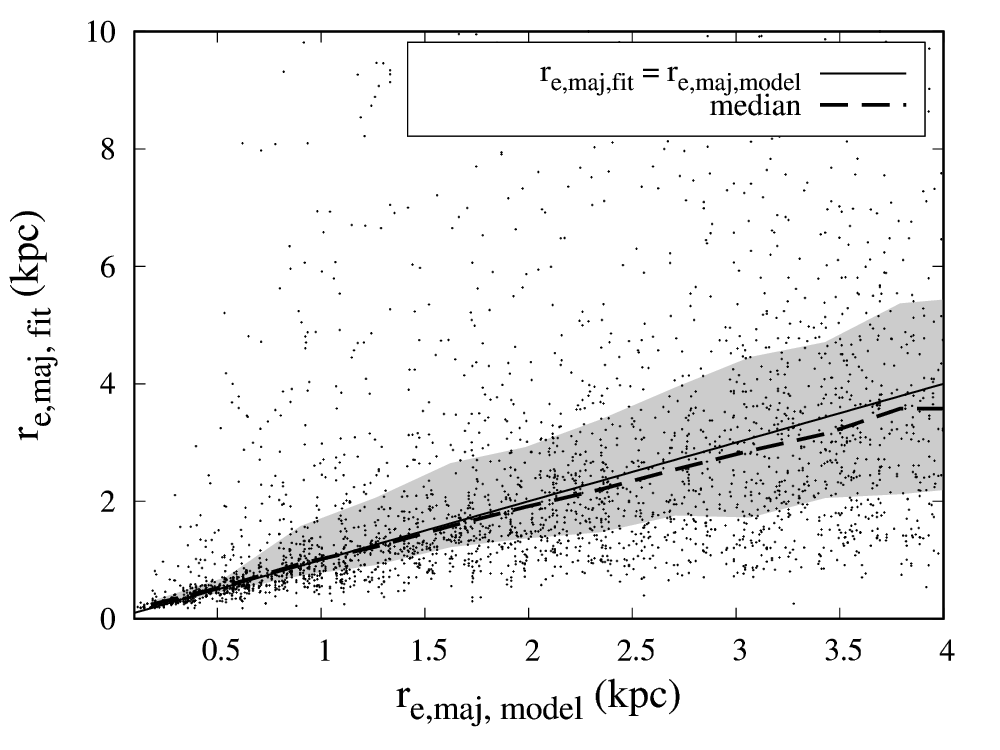}
\caption{
  {\it Top:} Sizes measured by {\sf GALFIT} ($r_{e,maj,fit}$) as a function 
  of input size ($r_{e,maj,model}$) in our simulation for ID1.
  The solid line shows $r_{e,maj,model}=r_{e,maj,fit}$.
  The dashed curve and gray shaded regions show the median 
  and 1$\sigma$ range of the $r_{e,maj,fit}$ at the given $r_{e,maj,model}$.
   {\it Bottom:} Similar to the {\it top} panel but for the stacked galaxy. 
}
\label{fig:galfiterror}
\end{figure}

%--------------------------
\begin{deluxetable*}{lcccccccc}
\tablecaption{Properties of the observed objects \label{tab:result}}
\tablecolumns{10}
\tablewidth{0pt}
\tablehead{
\colhead{ID} &
\colhead{$z_{\rm phot}$ } &
\colhead{$K_{tot}$\tablenotemark{a}} & \colhead{$K'_{tot, synth}\tablenotemark{b}$}
& \colhead{$K'_{tot, observed}$\tablenotemark{c}} & \colhead{$r_{e,maj}$\tablenotemark{d}}  
&  \colhead{ $M_{\star}$} \\
\colhead{} & \colhead{} & \colhead{(mag)}& \colhead{(mag)}  & \colhead{(mag)}  &  \colhead{(kpc)} & \colhead{($10^{11}~M_{\odot}$)} 
}
\startdata
   1   &   4.07 &  $22.47\pm0.05$&  $22.60^{+0.05}_{-0.05}$ & $23.26\pm0.06$ 
   & 	$0.92\pm0.31$ &  1.58\\ %SXDS1
   2   &   3.83 &  $22.54\pm0.05$&  $22.69^{+0.06}_{-0.05}$ & $23.43\pm0.08$ 
   &	$0.22\pm0.21$ & 1.09\\ %SXDS5
   3   &   3.70 &  $22.55\pm0.05$&  $22.65^{+0.05}_{-0.05}$ & $23.07\pm0.04$ 
   &	$0.63\pm0.18$ & 1.04\\ %SXDS2
   4   &   4.24 &  $22.61\pm0.05$&  $22.73^{+0.05}_{-0.05}$ & $22.85\pm0.03$ 
   &	$0.50\pm0.21$ & 1.83\\ %SXDS8
   5   &   4.09 &  $23.35\pm0.09$&  $23.38^{+0.10}_{-0.04}$ & $23.52\pm0.05$ 
   &	$1.70\pm0.71$ & 1.13\\ %SXDS6
   STACK   &   ...     & $22.54\pm0.03$ &   $22.67^{+0.03}_{-0.03}$ & $22.78\pm0.03$ 
   & $0.52\pm0.18$ & 1.38 \\
\enddata
\tablenotetext{a}{Kron magnitudes measured on the UKIDSS $K$-band images.}
\tablenotetext{b}{Expected $K'$-band total magnitudes from the SED fits.}
\tablenotetext{c}{Kron magnitudes measured on the IRCS-AO $K'$-band images.}
\tablenotetext{d}{Median and standard deviation of the $r_{e,maj}$ measured with fixed $n=0.5,1,2,3,4 \&5$. }
\end{deluxetable*}

%--------------------------
\begin{deluxetable}{lccc}
\tablecaption{{\sf GALFIT} fittings of ID5 with the IRCS-AO $K'$ and WFC3 $H$-band images \label{tab:k-candels}}
\tablecolumns{4}
\tablewidth{0pt}
\tablehead{
\colhead{Band} & \colhead{mag} &\colhead{$r_{e,maj}$} &\colhead{$n$}  \\
\colhead{} & \colhead{(mag)} & \colhead{(kpc)} & \colhead{}} 
\startdata
$K'$            &   $23.52\pm0.05$  & $1.70\pm0.71$ &$0.79_{0.5}^{2.99}$\\
$H$   &   $24.84\pm0.02$ & $1.19\pm0.03$ &$0.73\pm0.03$ \\
\enddata
\end{deluxetable}

%----------------------------------------------
\subsection{GALFIT fitting}
The sizes of our targets are measured by fitting S\'ersic profiles \citep{1968adga.book.....S} to their $K'$-band images 
using {\sf GALFIT} \citep{2002AJ....124..266P, 2010AJ....139.2097P}.  
{\sf GALFIT} fits two-dimensional analytical functions
convolved with a PSF to an observed galaxy image. 
Here we use a scale of $6.951$ kpc/arcsec 
in physical at $z=4$ for all the targets. 
We use the nearest star in the field of view or a star taken before and after the science exposures as the PSF reference star. 
We at first fit the S\'ersic models in ranges of effective radius 
$r_e=0.2-12$ kpc and S\'ersic index $n=0.5-10$.  
The fits are performed using an image cutout of 3.0 arcsec on a side for each object. 
The background values are estimated in an annulus between 
$2.9$ to $3.0$ arcsec from each object before S\'ersic model fittings. 
As an initial guess, we use total magnitudes measured by {\sf SExtractor} \citep{1996A&AS..117..393B}, 
$r_e =1$ kpc and $n=1.4$.
The results are not sensitive to this initial guess. 
In order to compare our results with \citet{2014ApJ...788...28V}, 
who measured galaxies sizes out to $z\sim3$, 
we here use the effective radius along the semi-major axis ($r_{e,maj}$). 

%--------------------------
\begin{figure*}
\centering
\includegraphics[width=175mm]{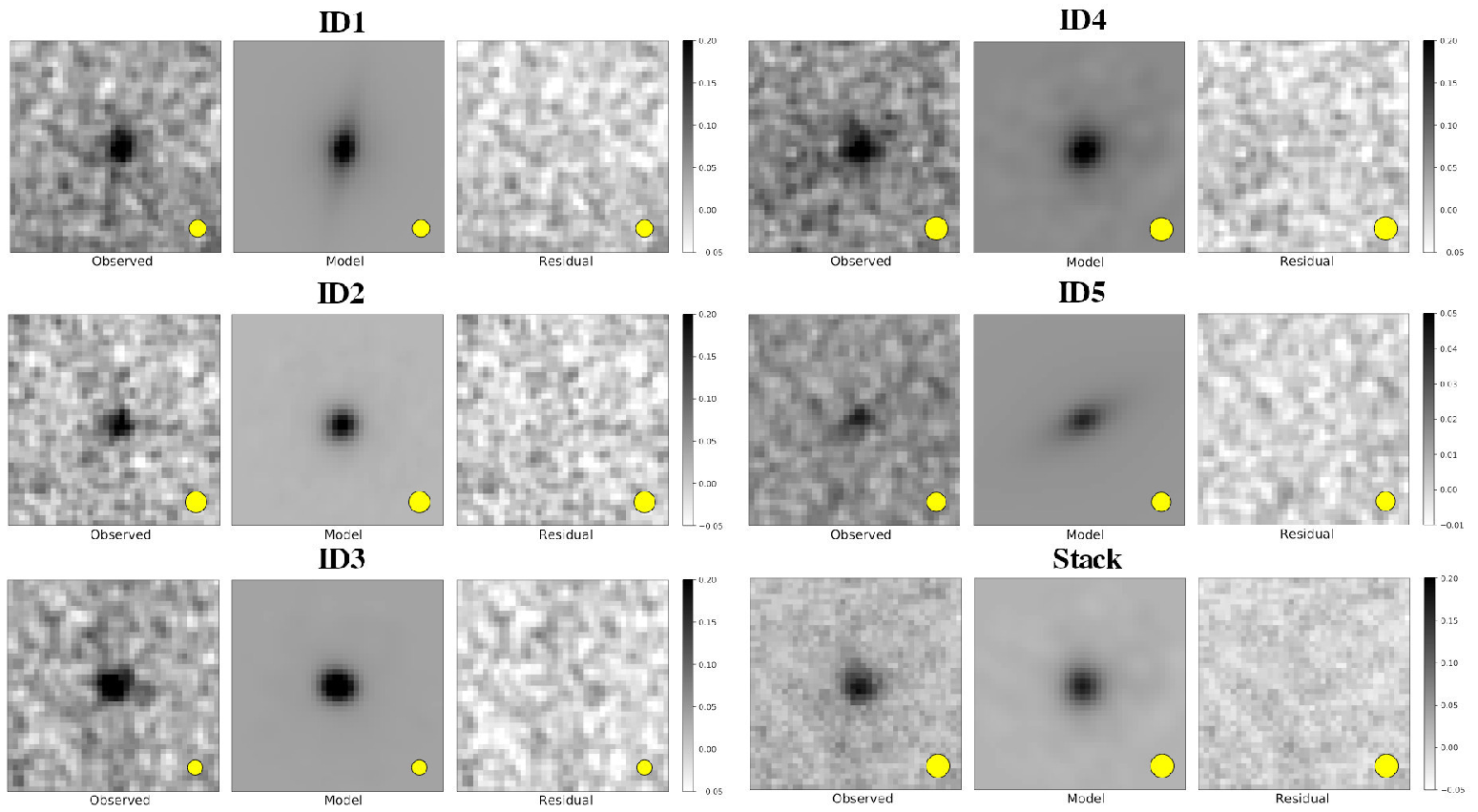}
\caption{
 The observed images and {\sf GALFIT} result for ID1 to ID5 and stacked galaxy.
  The observed images, best-fit S\'ersic models, and residuals are shown from left to right. 
  The sizes of the images are $2.1$ arcsec  $\times$ $2.1$ arcsec.
  The yellow filled circles show the FWHM of the PSF size on each image.
}
\label{fig:stamp}
\end{figure*}

%----------------------------------------------
\subsection{GALFIT fitting errors}
\label{sec:galfit_fitting_errors}

In \citet{2017MNRAS.469.2235K},  
the morphologies of galaxies at $z\approx3$ were studied
by using the deeper $K'$-band image taken with the same instrument with us 
and discussed the errors for {\sf GALFIT} fittings on those images.
We here discuss the possible errors in our size measurements following that work. 

Let us start with the limitation by PSF.
We are now studying the targets which can be hardly resolved 
even with our high resolution images. 
We should note that our results can be just an upper limit 
since the reduced $\chi^2$ values of S\'ersic model fitting 
only marginally ($\Delta \chi^2\sim 0.01$) improves from that with PSF model fitting. 
In addition, the fits with models of different S\'ersic indices $n=0.5\sim5$ 
are equally good. 
Then we adopt the median of $r_{e,maj}$ 
of GALFIT fitting with $n=0.5,1,2,3,4~\&~5$ 
as the best-fit values. 

In addition, there can be errors originated in a little PSF inconsistency.
We ideally need to evaluate the PSF at the positions of the targets,
but that is in practice difficult.  
We use a single PSF reference star either
within the field of view or taken before/after the science exposures. 
Even though the target and PSF reference stars are taken in a same frame, 
as shown in Table \ref{tab:targetdiscription}, 
the distance between the tip-tilt star and the target, 
and that between the tip-tilt star and the PSF reference star are not the same.
In case of our targets, we expect the PSF difference of $\lesssim0.03$ arcsec 
according to the performance of AO188\footnote{https://www.subarutelescope.org/Observing/Instruments/AO\\/performance.html}.
However since the size of galaxies at $z\sim4$ is very small, this may not be negligible.
The separation between the PSF reference star and the tip-tilt star is always smaller
than the separation between the target and the tip-tile star, i.e., the PSF we use
in the fits is likely smaller than the real PSF at the object position.
This leads us to over-estimate the size.  Thus, our estimates are likely conservative.
\citet{2017MNRAS.469.2235K} reported that, 
this level of PSF inconsistency does not affect the measured sizes but 
on the other hand, it significantly affects the measured S\'ersic indices, 
which we do not discuss in this paper.

Next, we test the accuracy of the {\sf GALFIT} measurement 
by generating mock galaxy images following \citet{2017MNRAS.469.2235K}.
We investigate the typical fitting errors by inserting artificial objects 
on the sky of the observed image, measuring the sizes of them 
and comparing the input and output structural parameters.
We here use the coadd image of ID1 as the representative case of our sample. 
We generate artificial sources over a range of parameters; 
$K'=22.6$ ($\approx K'_{tot, synth}$  of ID1), 
$r_{e,maj}<4$ kpc, S\'ersic indices $n=0-8$, and various axis ratios and position angles.  
They are convolved with the PSF reference star for ID1 
and added to the sky of ID1 image.  
By repeating this simulation, we find that
the median and standard deviation of the measured total magnitude 
are $23.0\pm0.6$ in case $r_{e,maj,model}=0.5\sim2$ kpc, 
which is consistent with the observed $K'$-band total magnitude of ID1.
In other words, the simulation reproduces the missing flux in the observation,
suggesting that our simulation is reasonably realistic. 
We show the $r_{e,maj}$ of the mock galaxies ($r_{e,maj,model}$) 
v.s. those measured on them with {\sf GALFIT} ($r_{e,maj,fit}$) in Fig. \ref{fig:galfiterror}. 
Naively, we expect that small ($\la 1$ kpc) objects are overestimated the sizes from the input sizes
due to the limited resolution while large objects are underestimated 
since their outer profiles are buried in noise. 
We can see this tendency weakly. 
The standard deviation of S\'ersic indices is $\sigma(n)=2.3$ (plot not shown). 
This again suggests that the S\'ersic indices are hardly constrained with our data.
Though we input various axis ratios, the measured axis ratios tend to be lower than 0.4, 
more asymmetric models are favored as the best-fit models. 
This may be caused by asymmetric distribution of noisy pixels around the source, 
since this tendency is soften at the depth of the stacked image.
The best-fit models of ID1 and ID5 in Fig. \ref{fig:stamp} look elongated 
however it is not clear they are real signatures. 

Finally, we compare the $r_{e,maj}$ measured
with {\it HST}/WFC3 $H$-band image and our $K'$-band images. 
Among our sample, only ID5 is within the Cosmic Assembly 
Near-IR Deep Extragalactic Legacy Survey (CANDELS;
\citealt{2011ApJS..197...35G,2011ApJS..197...36K}). 
We summarize the comparison in Table. \ref{tab:k-candels}.
They should not necessarily be the same as our result 
due to the wavelength difference but are a good comparison. 
The size estimates on these images are broadly consistent with each other, but
our size estimate is slightly larger as expected from the simulation above.
This may also imply that they show no strong rest-frame UV to optical color gradient
due to age and/or metallicity gradient of the stellar population as well as attenuation by dust. 
The stellar population of these galaxies may be relatively simple. 
On the other hand, uncertainty in S\'ersic indices is large for $K'$, 
which is again consistent with the above indications.

Taken all the tests together, there is a small bias in our size measurements
for individual objects in the sense that we tend to over-estimate the sizes
by $\sim30$ \%.  
We do not account for this bias just to be conservative. 
Our estimates can thus be considered as reasonable upper limits.

%----------------------------------------------
\subsection{Stacking analysis}

We stack our targets to gain S/N and measure their average size.
We exclude ID5 from the stacking because it is relatively fainter than the others.  
We smoothed  the single exposure images  of ID1 to ID4 
to a common seeing of $0.23$ arcsec (that of ID4) by convolving with a Gaussian 
and then performed median stacking of them.  
The total $K'$-band magnitude measured on the stacked image shows only a small amount
of missing flux (10\%, Fig.~\ref{fig:magauto}).

We repeat the same {\sf GALFIT} simulation 
using the stacked image and PSF reference star for ID4 (Fig. \ref{fig:galfiterror}, {\it bottom}). 
Similar to the individual galaxies, the reduced $\chi^2$ values of S\'ersic model fitting 
only marginally improves from that with the PSF model fitting. 
The $r_{e,maj}$ errors are reduced greatly from the simulation for ID1. 
The bias in the size measurement marginally changes depending on the PSF adopted. 
This gives us a confidence on the measured sizes of QGs on our stacked image.

%----------------------------------------------
\section{Results}

The results of {\sf GALFIT} fitting are shown
in  Fig. \ref{fig:stamp} and summarized in Table \ref{tab:result}. 
The $r_{e,maj}$ of our targets range from $0.2-1.7$ kpc with
the median and standard deviation being 0.6 kpc and 0.6 kpc, respectively.
Our results indicate that massive QGs at $z\sim4$ are indeed compact.
As discussed above, the individual size estimates may suffer from the flux
incompleteness (we are missing diffuse light), but we obtain a consistent result
for the stacked galaxy; the $r_{e,maj}$ measured on the stack is $0.52\pm0.18$ kpc,
providing further support for the compact sizes.
We also note that the possible systematic errors from the PSF inconsistency 
are not included in our size estimate errors, 
however, as we mentioned above, it may not affect them significantly.

\begin{figure}
\centering
\includegraphics[width=85mm]{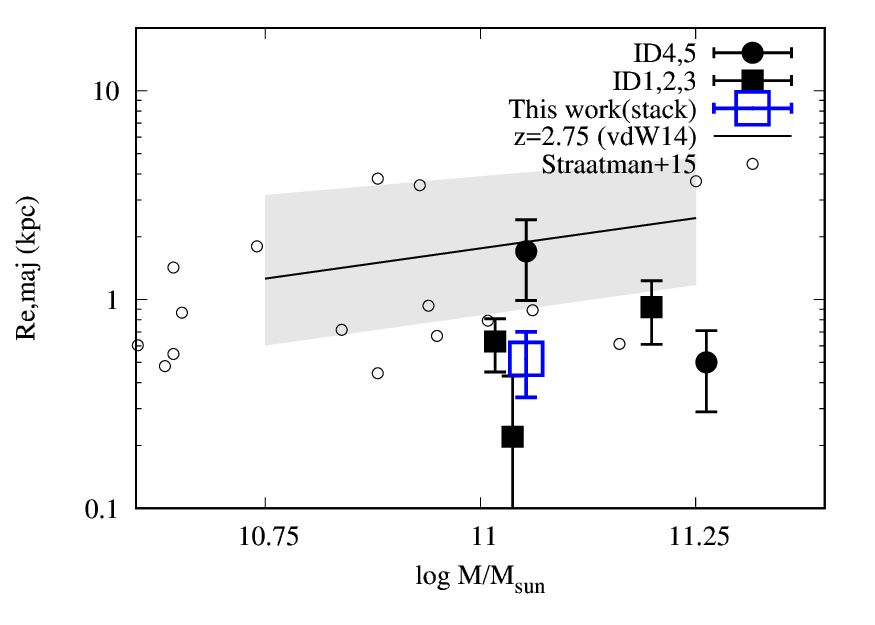}
\caption{
  Stellar mass v.s. $r_{e,maj}$. 
  The filled squares, filled circles and  blue open square 
  shows ID1-3, ID4-5 and the stack of ID 1-4, respectively. 
  The solid line and the shaded area show the size-stellar mass relation for QGs at $z=2.75$
  %and its 68\% range
  in \citet{2014ApJ...788...28V}. 
  The open circles show QGs at $z\sim3.7$ measured the sizes 
  at rest-frame UV in \citet{2015ApJ...808L..29S}.
  \label{fig:masssize} }
\end{figure}

Figure \ref{fig:masssize} shows the stellar mass v.s. $r_{e,maj}$ diagram
of the QGs at $z\sim4$.
For comparison, we plot the size-stellar mass relation of
QGs at $z=2.75$ measured at  rest-frame optical \citep{2014ApJ...788...28V} 
and at $z\sim3.7$ measured at rest-frame UV 
(\citet{2015ApJ...808L..29S} using the catalog in \citet{2014ApJ...783L..14S} described above). 
Both studies select QGs with the photometric redshifts and $UVJ$ colors,  
and measure the size on {\it HST}/WFC3 $H$-band images from CANDELS. 
The QGs at $z\sim4$ are below the size-stellar
mass relation of QGs at $z=2.75$, suggesting that they have 
the physical sizes smaller than lower redshift ones.  
The size measured on the stack shown with the open square confirms this trend.
The QGs at $z\sim3.7$ have a somewhat large dispersion in size, but our targets have
consistent sizes with some of their smallest objects.  
There are a few objects with a large size of $r_{e,maj}\sim4$ kpc among QGs at $z\sim3.7$
which are more consistent with the typical sizes 
of SFGs at $z=2.75$ \citep{2014ApJ...788...28V}.  
This might indicate the contamination 
of SFGs in their $UVJ$-selected QGs.

\begin{figure}
\centering
\includegraphics[width=85mm]{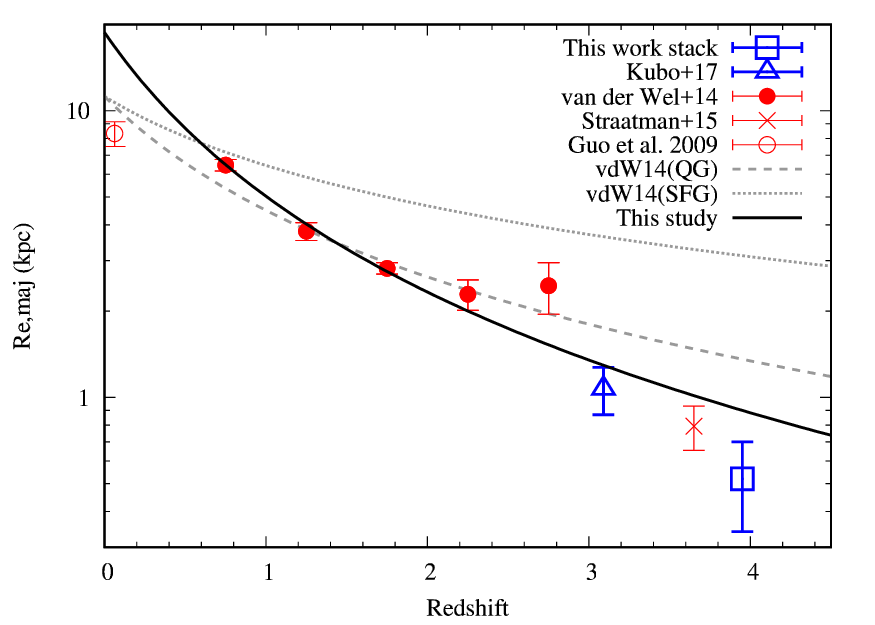}
\caption{
  Size evolution of  QGs with the stellar mass $10^{11}~M_{\odot}\leq M_{\star}\leq 10^{11.5}~M_{\odot}$ at up to $z=4$. 
  The blue open square shows the stack of QGs at $z\sim4$ in this study.
  The red open circle, filled circles and cross show the median 
  $r_{e,maj}$ of QGs at $z=0$ from \citet{2009MNRAS.398.1129G}, 
  at $0.75\leq z\leq2.75$ from  \citet{2014ApJ...788...28V} 
  and $z\sim 3.7$ from  \citet{2015ApJ...808L..29S}, respectively.
  The error bar of \citet{2009MNRAS.398.1129G} shows the $\sim0.1$ dex 
  difference between \citet{2003MNRAS.343..978S} and \citet{2009MNRAS.398.1129G}. 
  The blue open triangle shows the QG with $M_{\star}\approx2\times10^{11}~M_{\odot}$ 
  at $z=3.1$ in \citet{2017MNRAS.469.2235K}.  
  The black solid curve shows $r_{e,maj}=A(1+z)^{\beta}$ fit in this study.  
  The gray dashed and dotted lines show those for QGs and SFGs 
  %with $10^{11}~M_{\odot}\leq M_{\star}\leq 10^{11.5}~M_{\odot}$
   in \citet{2014ApJ...788...28V}, respectively.
  \label{fig:sizeredshift}}
\end{figure}

In Fig. \ref{fig:sizeredshift}, we show the rest-frame optical size-redshift relation 
of galaxies with $10^{11}~M_{\odot}\leq M_{\star}\leq 10^{11.5}~M_{\odot}$.
The $r_{e,maj}$ at $z=0$, $0.75\leq z\leq2.75$ and $z\sim 3.7$ 
are the median sizes of QGs with $10^{11}~M_{\odot}\leq M_{\star}\leq 10^{11.5}~M_{\odot}$ 
from \citet{2009MNRAS.398.1129G},  \citet{2014ApJ...788...28V} 
and \citet{2015ApJ...808L..29S}, respectively.
The $z=3.1$ point is from \citet{2017MNRAS.469.2235K} 
who measured the size of a QG with $M_{\star}\approx 2\times10^{11}~M_{\odot}$ 
in a protocluster at the $K'$-band using IRCS/AO188.
Our stacked galaxy is shown as the open square.
% (individual objects are not shown for the sake of clarity).  
We extend the size-redshift relation of QGs out to $z\sim4$ for the first time.  
The figure shows that the sizes of massive QGs continuously decreases with redshift up to $z=4$,
an order of magnitude size evolution between $z=4$ and 0.  
This is a surprisingly strong evolution. 
Note that the size-stellar mass relation in \citet{2014ApJ...788...28V} show an upturn at $z=2.75$. 
This could be caused by the contamination of SFGs. 
In their $UVJ$ color diagram, 
the dispersions of color sequences of QGs and SFGs increase with redshift 
due to the observational errors and maybe the change of the SEDs of galaxies. 
Then it is expected that contaminants in galaxies selected as QGs increase with redshift. 
\citet{2015ApJ...808L..29S} also uses rest-frame $UVJ$ color selection 
but since they did not use the sample near the border of selection criterion,  
such contaminants may be reduced in their sample.

The size-redshift relation is often parameterized in a form $r_e/{\rm kpc}=A(1+z)^{\beta}$. 
\citet{2014ApJ...788...28V} find $A=11.2_{-2.1}^{+2.6}$ and $\beta=-1.32\pm0.21$ 
for QGs with $10^{11}~M_{\odot}\leq M_{\star}\leq 10^{11.5}~M_{\odot}$ 
(dashed line in Fig. \ref{fig:sizeredshift}).
Adding the results at $z>3$ and fitting at $0.75\leq z \leq 4$, 
we find $A=18.8\pm3.0$ and $\beta=-1.9\pm0.2$ (solid line), 
though it is hard to fit the whole redshift range with this form. 
Our results support a stronger size evolution of QGs compared to SFGs
with $\beta \sim-1$  (e.g., \citealt{2014ApJ...788...28V,2015ApJS..219...15S,2015ApJ...808L..29S}) up to $z=4$.

%--------------------------

%--------------------------

%------------------------------------------------------
\section{Discussion}

In this study, we measure the size of massive QGs 
at $z\sim4$ in the rest-frame optical wavelength
for the first time based on the AO-assisted imaging using a ground-based telescope.  
There are a few possible uncertainties in our results. 

One is contamination of AGNs 
which could make galaxies look compact.
However, as shown in Fig.~\ref{fig:individual_seds}, 
the overall SEDs of our targets are dominated by evolved
stellar populations as indicated by the strong Balmer break, 
which suggests that the continuum is dominated by stars.  
Thus, the AGN contamination, if any, is unlikely to significantly alter our results.
Our targets are not detected in X-ray \citep{2008ApJS..179..124U} or MIR \citep{2007sptz.prop40021D}.
Although only very active AGNs are detectable at the depth of the data at $z\sim4$, 
this adds further support for no significant AGN contamination.

There is another question of the quiescence of our targets.
Although the SED fits indicate that these galaxies are not actively forming stars, 
their quiescence should be further confirmed by other means. 
\citet{2017arXiv170302207G} detected 
significant far-IR fluxes from $BzK$ and $UVJ$-selected QGs,
suggesting that the optical-nearIR selection does not always give a clean sample of QGs.
Multi-wavelength follow-up observations of our targets are essential to fully confirm their quiescence.
Efforts in this direction are underway.

Although we should further address these possible uncertainties in the future, 
it is interesting to discuss the origin and evolution 
of these extremely compact massive QGs at $z\sim4$. 
In this section,  we first discuss the extremely high stellar mass surface density 
of them and then focus on their size evolution on the evolving stellar mass track.

\subsection{Extremely high stellar mass surface density}

%--------------------------
\begin{figure}
\centering
\includegraphics[width=85mm]{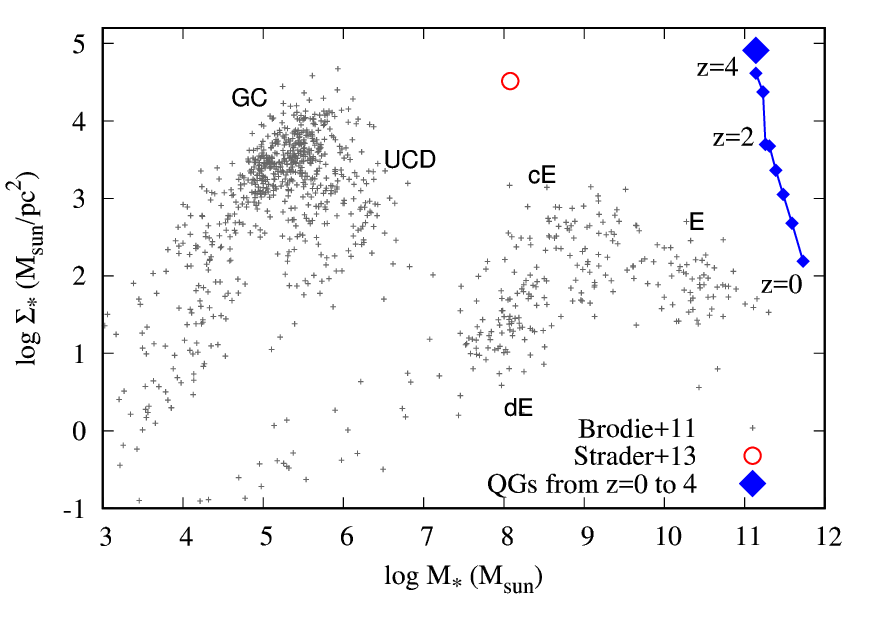}
\caption{
 Surface stellar mass densities within effective radii v.s. stellar mass. 
 The large blue diamond shows massive QGs at $z=4$. 
 We also show its evolution track found in \S 6.2. 
 The crosses show dispersion-supported systems in local Universe 
 from \citet{2011AJ....142..199B}
 (GC=globular cluster; cE=compact elliptical; E=early-type galaxy; dE=dwarf elliptical).
 The red open circle shows the densest UCD reported in \citet{2013ApJ...775L...6S}.
   \label{fig:surfacedensity}  }
\end{figure}

It has been know that massive QGs at high redshift 
have extremely high stellar mass surface densities
(e.g., \citealt{2008ApJ...677L...5V}). 
We compare the mean stellar mass surface densities 
within the effective radii of massive QGs at $z=4$ 
and dispersion-supported stellar systems in the local Universe 
\citep{2011AJ....142..199B} in Fig. \ref{fig:surfacedensity}. 
\citet{2011AJ....142..199B} is originally given in $V$-band luminosity. 
We convert the $V$-band luminosity into stellar mass adopting $M_{\star}/L_V=3$ 
which is in case of a simple stellar population model with the age of $\sim10$ Gyr 
adopting \citet{2003PASP..115..763C} IMF.
Note that $M_{\star}/L_V$ can depend on the object type. 
We also show the densest ultra compact dwarf (UCD) reported in \citet{2013ApJ...775L...6S} 
using its stellar mass from \citet{2014Natur.513..398S}.

It is interesting that high-z QGs and Globular clusters (GCs), 
consisting of the oldest stars of the Milky Way and thought to form at high redshifts, 
both have extremely high stellar mass surface densities, 
even though their typical mass differ by several orders of magnitude.  
\citet{2017ApJ...843...78J} and \citet{2017MNRAS.467.4304V} 
find very low mass ($M_{\star}=$ a few $10^6~M_{\odot}$) 
extremely dense galaxies at $z=2-6$ with strong lensing. 
Though they are more massive than GCs, 
they imply that such ultra dense objects are commonly formed at high redshift. 
Given the high density and high gas fraction in the early Universe, 
we naturally expect that gas rich major mergers 
are one of the channels to form such extremely compact objects.  
In addition, cosmological numerical simulations predict that 
high-z galaxies are fed by streams of smooth gas and merging clumps from the cosmic web,  
and then they are settled into violent disc instabilities and end up with dense objects 
from dissipative compaction of gas and subsequent starburst 
\citep{2014MNRAS.438.1870D, 2015MNRAS.450.2327Z}.

We remark that at $z > 4$, dusty SFGs (DSFGs) 
display very compact far-IR emitting regions 
that locate the ongoing starburst and establish a good proxy for the subsequent stellar remnant 
(e.g., \citealt{2015ApJ...810..133I, 2017arXiv170904191O, 2018ApJ...856..121G}). 
In a sample of six DSFGs at $z \sim 4.5$ with evidence of minor mergers, 
\citet{2018ApJ...856..121G} measured a median stellar mass 
of $\log (M_{\star}/M_{\odot}) = 10.49 \pm 0.32$ and far-IR sizes of $r_{e} = 0.70 \pm 0.29$ kpc. 
They expect the starburst to be completed in $\sim 50$ Myr, faster than the anticipated timescale for the observed mergers of $\sim 500$ Myr. 
Massive QGs at $z\sim4$ studied here may have stopped star formation earlier ($z>5$) 
than these DSFGs, however, they present the capability of quickly building up and quenching of 
massive stellar cores at such high redshift. 
Further detailed studies of DSFGs with ALMA are awaited. 

We finally quote \citet{2010MNRAS.401L..19H} 
which reports that the maximum stellar surface 
densities of GCs and high-z compact QGs 
are at the global stellar mass surface density limit 
regardless of their masses and propose that it is limited 
by feedback from young massive stars when star formation reaches the Eddington limit. 
Their results also imply that the densest objects 
are formed in the extreme situation which may be only achievable in the early Universe. 

\begin{figure}
\centering
\includegraphics[width=87mm]{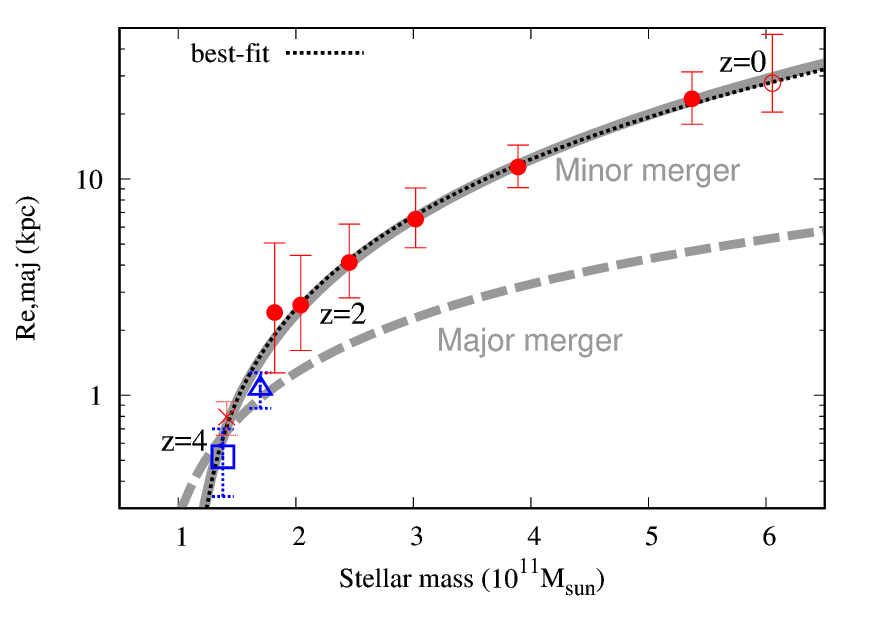}
\caption{
 Size-stellar mass growth from massive QGs at $z=4$ taking the stellar mass evolution 
 into account based on \citet{2014ApJ...794...65M}. 
 The $r_{e,maj}$ at each point are that extrapolated from \citet{2014ApJ...788...28V} (red filled circles), 
 median and the 25-75\% interval of the $r_{e,maj}$ 
 of galaxies with $M_{\star}=10^{11.8}~M_{\odot}$ 
 at $z=0$ from \citet{2009MNRAS.398.1129G} (red open circle), 
 and observed values (others). 
 The black dotted curve shows the best-fit curve. 
 The gray solid and dashed curves show the toy models of size-stellar mass growth 
 in cases of minor mergers ($r_{e,maj}\propto M_{\star}^2$) 
 and major mergers ($r_{e,maj}\propto M_{\star}$), respectively.
  \label{fig:size-mass-ccd}}
\end{figure}

\subsection{Size evolution on the evolving stellar mass track}

The size-redshift relation of massive QGs in Fig. \ref{fig:sizeredshift} is 
measured at a fixed stellar mass. 
Galaxies grow with time and evolving cumulative number density
determined in the semi-empirical approach using abundance matching 
has been used to find the progenitors of particular descendants  
(e.g., \citealt{2013ApJ...777L..10B}).
\citet{2014ApJ...794...65M} tracks the progenitors 
of ultra-massive galaxies today
($M_{\star}\approx10^{11.8}~M_{\odot}$) by this method 
and find their stellar mass evolution as a function of redshift, $\log (M_{\star}/M_{\odot}) = A + Bz + Cz^2$ 
where $A=11.801\pm0.038$, $B=-0.304\pm0.054$ and $C=0.039\pm0.014$. 
This relation is based on the observations at $z<3$ 
but if we extrapolate it to $z\sim4$,  we find that 
massive QGs at $z\sim4$ in this study are on this evolutionary track,
i.e., they plausibly evolve into ultra-massive galaxies today.

\begin{figure}
\centering
\includegraphics[width=85mm]{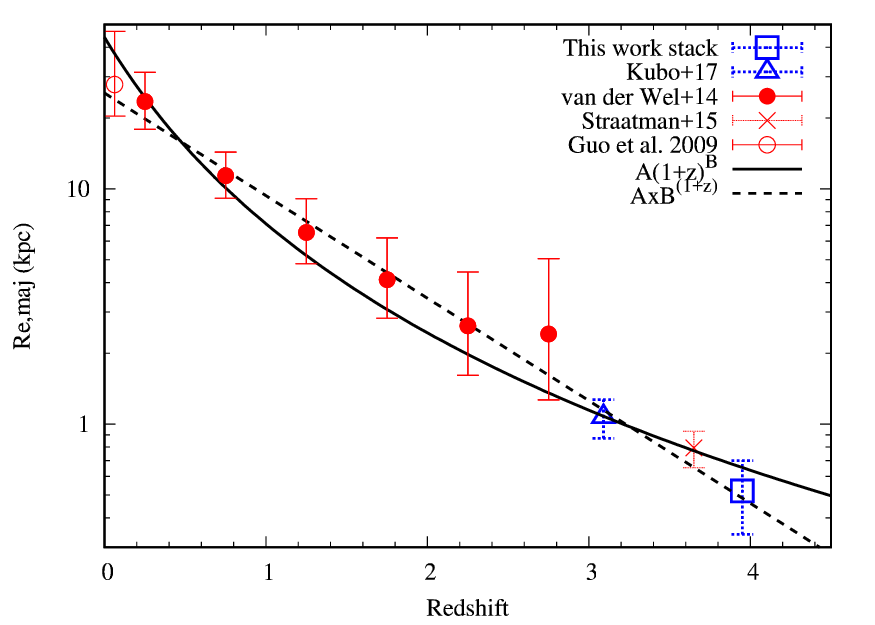}
\includegraphics[width=85mm]{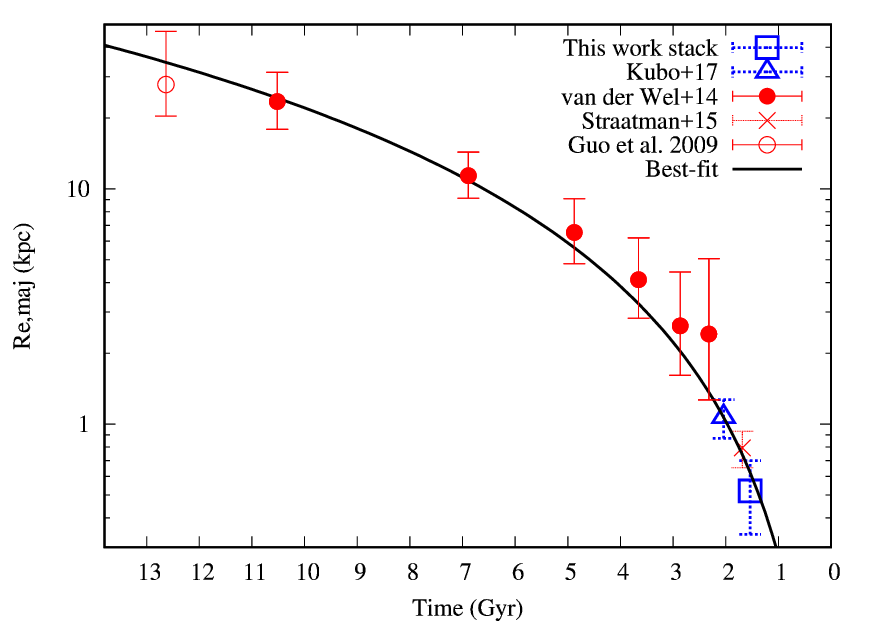}
\caption{
  {\it Top:} Size-redshift relation of massive QGs taking the stellar mass evolution into account.
  The data points are the same as those in Fig. \ref{fig:size-mass-ccd}.
  The black solid curve and dashed line show the best-fit curves 
  in forms of $r_{e,maj}/{\rm kpc}=A\times(1+z)^{B}$ and $r_{e,maj}/{\rm kpc}=A\times B^{-(1+z)}$, respectively.
  {\it Bottom:} Similar to the {\it top} panel but scaled in cosmic time 
 and  the black solid curve shows the best-fit curve in a form of $\log(r_{e,maj}/{\rm kpc})= A+B \log(t/\rm Gyr)$.
  \label{fig:sizeredshift-constatnt}}
\end{figure}

Taking the stellar mass evolution into account 
using the stellar mass-redshift relation in \citet{2014ApJ...794...65M}, 
we show the size-stellar mass evolution from massive QGs at $z=4$ in Fig. \ref{fig:size-mass-ccd}. 
 The $r_{e,maj}$ are median and the 25-75\% interval of the $r_{e,maj}$ 
 of galaxies with $M_{\star}=10^{11.8}~M_{\odot}$ 
 from \citet{2009MNRAS.398.1129G} at $z=0$, 
 and extrapolated from the size-stellar mass relation 
 at each redshift in \citet{2014ApJ...788...28V} at $0.25\leq z\leq 2.75$. 
The $r_{e, maj}$ at $z>3$ are the observed values  
since their stellar masses are on the stellar mass-redshift relation of \citet{2014ApJ...794...65M}. 
The point at $z=3.1$ \citep{2017MNRAS.469.2235K} is not included in the fit, 
this is shown just for a reference. 
The {\it top} and {\it bottom} panels of Fig. \ref{fig:sizeredshift-constatnt} 
show the size evolution as functions of redshift and cosmic time, respectively. 
The size-redshift relation is fitted in a form of $r_{e,maj}/{\rm kpc}=A\times(1+z)^{B}$ 
where $A=44.1\pm6.1$ and $B=-2.6\pm0.2$ or $r_{e,maj}/{\rm kpc}=A\times B^{-(1+z)}$ 
where $A=69.7\pm7.7$ and $B=2.7\pm0.1$. 
If we fit the size-time relation in a form of $\log(r_{e,maj}/{\rm kpc})= A+B \log(t/\rm Gyr)$,  
we obtain  $A=-0.56\pm0.07$ and $B=1.91\pm0.09$. 
We find that they grow in size by a factor $\sim10$ in the first few Gyr
and finally acquire the size $\sim30$ times larger than that of a massive QG at $z=4$ by $z=0$. 
They have also evolved significantly in the stellar mass surface density (Fig. \ref{fig:surfacedensity}). 

In order to constrain physical processes driving this rapid evolution, 
we compare the size-stellar mass growth
to the two toy models shown in Fig. \ref{fig:size-mass-ccd}. 
We show size-stellar mass growth models 
via minor mergers (gray solid curve) and major mergers (gray dashed curve), 
which follow $r\propto M^2$ and $r\propto M$, 
respectively \citep{2009ApJ...697.1290B,2009ApJ...699L.178N}.
The observed size-stellar mass growth closely follows that of minor mergers: 
it is fitted in a form $r_{e,maj}/{\rm kpc}=A\times (M_{\star}/10^{11}~M_{\odot})^B + C$ 
where $A=1.0\pm0.4$, $B=1.9\pm0.2$ and $C=-1.3\pm0.6$ (black dotted curve).  
Similarly, \citet{2010ApJ...709.1018V} evaluates
the size-stellar mass evolution of massive galaxies 
with $M_{\star}\approx10^{11.45}~M_{\odot}$ at $z=0$
taking the stellar mass evolution into account by constant number density method. 
Although the samples and methods are different, 
they report $r_e\propto M^{2.08}$ evolution, similar to our result.
Our result also agree with the prediction of the stellar mass and size growth 
of massive-end QGs with $M_{\star}\approx10^{11.8}~M_{\odot}$ 
in \citet{2018MNRAS.474.3976G} based on the IllustrisTNG simulation.
Taking all these results together, we conclude that 
the evolution of massive-end galaxies from $z=4$ 
is likely to be driven by minor mergers. 

Note that lower mass galaxies may not necessarily 
follow the size growth found in this study. 
The mass dependent evolution has been predicted in cosmological numerical simulations. 
E.g., more moderate size growth of lower mass galaxies is predicted in  
\citet{2018MNRAS.474.3976G}. 
The continual addition of massive galaxies to the quiescent population, 
so called progenitor bias may also contribute 
to the observed size growth (e.g., \citealt{2013ApJ...773..112C,2013ApJ...777..125P}) 
though it alone may not be sufficient \citep{2015ApJ...799..206B}.
Several studies reported that the observed merger rate 
is not capable for the size growth of high-z compact ellipticals
\citep{2011ApJ...738L..25W,2012ApJ...746..162N,2016ApJ...830...89M}
but on the other hand, in situ star formation in satellites 
before mergers can push up the size-growth 
amount via minor mergers \citep{2016ApJ...816...87M}. 
It can also happen that the environment of the most massive galaxies is special. 
Massive compact elliptical at $z=3.09$ cited from \citet{2017MNRAS.469.2235K} 
is in a dense group of massive galaxies capable for the ten times size growth at least. 
Further studies of not only compact massive quiescent galaxies themselves 
but also their environment are needed to understand 
what physical processes govern the size-stellar mass growth.

\section{Conclusion}

We have measured the rest-frame optical sizes of massive galaxies 
with suppressed star formation at $z\sim4$ 
with IRCS and AO188 on the Subaru telescope.
Although our measurements on individual galaxies are noisy, the more robust size measurements on the stacked object reveals that they have smaller physical sizes compared to lower redshift ones. 
This is the first measurement of the rest-frame optical sizes of QGs at $z\sim4$. 
Their mean stellar mass surface density is similar to those of GCs, the densest objects of the Universe, 
although their masses differ by several orders of magnitude.  
This implies that the origin of the densest galaxies 
are due to the high density and high gas fraction in the early Universe.  
If we take the stellar mass evolution  into account,  
they plausibly evolve into the most massive galaxies today 
and their stellar mass-size evolution is consistent with a scenario in which minor dry mergers drive the size growth. 

We have shown that massive QGs at $z\sim4$ are compact, 
but we have pushed the ability of current facilities close to the limit. 
Deeper and higher resolution imaging  
at $> 2~\mu$m with AO on ground based large(r) telescopes 
and James Webb Space Telescope ({\it JWST}) are needed to make a leap from here. 

%------------------------------------------------------
%\section{Conclusions}
%In this study, we selected the candidate QGs at $z\sim4$ 
%from deep and wide multi-band survey in SXDS field 
%and estimated their sizes by using the AO-assisted $K'$-band images. 
%This is the first time to measure the sizes of QGs at $z\sim4$ properly in rest-frame optical. 
%We show that QGs continues to become compact with redshift up to $z\sim4$. 

\acknowledgments
K.Y. was supported by JSPS KAKENHI Grant No. JP16K17659 and JP18K13578.
M.T. acknowledges support by JSPS KAKENHI Grant No. 15K17617.
MS and S.T. acknowledge support from the European Research Council (ERC) Consolidator Grant funding scheme (project ConTExt, grant number 648179). The Cosmic Dawn Center is funded by the Danish National Research Foundation.
This work is based on data collected at the Subaru Telescope, which is operated by
the National Astronomical Observatory of Japan.
We thank the anonymouns referee for the useful report, which helped improve the paper.

\end{document}